\documentclass[final,5p,times,twocolumn]{elsarticle}

\usepackage{amssymb}
\usepackage{siunitx}
\usepackage{hyperref}
\usepackage[version=4]{mhchem}
\usepackage[labelformat=simple]{subcaption}
\usepackage{tikz}
\usepackage{textgreek}

\usepackage{pgfplots}
	\pgfplotsset{compat=1.16}
	\pgfdeclarelayer{background}
	\pgfdeclarelayer{foreground}
	\pgfsetlayers{background,main,foreground}
	\usepgfplotslibrary{fillbetween}
	\usepgfplotslibrary{dateplot}

\journal{NIM A}

\begin{document}

\begin{frontmatter}



  \title{Construction and characterization of a seven-chip GridPix X-ray detector for solar axion searches}

   \author[label1]{Klaus Desch}
   \author[label1]{Jochen Kaminski}
   \author[label1,label2]{Christoph Krieger}
  \cortext[cor1]{Corresponding authors:}
  \author[label1]{Tobias Schiffer\corref{cor1}}
  \ead{schiffer@physik.uni-bonn.de}
  \author[label1]{Sebastian Schmidt\corref{cor1}}
   \ead{phd@vindaar.de}

    \affiliation[label1]{organization={Physikalisches Institut der Universität Bonn},
    addressline={Nußallee 12},
    city={Bonn},
    postcode={53115},
    state={NW},
    country={Germany}}

    \affiliation[label2]{organization={Institut für Experimentalphysik der Universität Hamburg},
    addressline={Luruper Chaussee 149},
    city={Hamburg},
    postcode={22761},
    state={HH},
    country={Germany}}

  \begin{abstract}
  In the scope of solar axion searches, detectors which are able to
measure low energetic X-rays with high efficiency are required. For
this purpose a detector based on the GridPix technology was built for the CAST experiment at CERN. The
GridPix is a pixelised readout ASIC (Timepix) with a Micromegas-like
gas amplification stage (grid) built photolithographically on top.  In
order to reduce the detector´s background level, several hardware and
software vetoes were implemented.

  Hardware-wise, these vetoes consist of a ring of six GridPixes around
a central GridPix, a readout of the induced grid signal, and two
scintillators. On the software side, multiple approaches to
distinguish between background events and X-ray photons are
implemented. Here, also the hardware features, like the six
surrounding GridPixes, are used.

The new detector was tested in a long ($3500\,\text{h}$) background data taking campaign. The performance of the new vetoes was evaluated. The detector performance itself, for low energetic X-rays, was also evaluated with a variable X-ray generator using eight different energies from 0 to $10\,\text{keV}$. The efficiency for very low energetic
X-rays and the energy resolution was determined.
  \end{abstract}


  \begin{highlights}
  \item Solar axions
  \item Ultra-low background detector
  \end{highlights}


\end{frontmatter}


\section{Introduction}
\label{Intro}

\begin{figure}[htb]
    \centering
    \begin{tikzpicture}[font=\footnotesize]
	    \begin{axis}[
            grid = major,
            xlabel={Energy [keV]},
            ylabel={Flux $[10^{20}\,\text{keV}^{-1}\text{m}^{-2}\text{yr}^{-1}]$},
            ytick scale label code/.code={},
			xmin=0.01, xmax=15,
			ymin = 0,
			scaled x ticks=false,
			width=0.95\linewidth,
			height=0.75\linewidth,
            legend style={at={(0.97,0.97)},anchor=north east,cells={align=left}}
			]
        \addplot[red, line width=0.5pt] table[skip first n=1, x expr =(\thisrowno{0}/1000), y expr =(\thisrowno{7}*100), col sep=space] {axion_flux.csv};
        \addlegendentry{Primakoff flux $\cdot 100$}
		\addplot[blue, line width=0.5pt] table[skip first n=1, x expr =(\thisrowno{0}/1000), y expr =(\thisrowno{9}), col sep=space] {axion_flux.csv};
        \addlegendentry{Total flux}
	\end{axis}
    \end{tikzpicture}
  \caption{\label{fig:axion_spectrum}Comparison of the axion spectrum for different possible production mechanisms. This figure uses \(g_\text{ae} = \num{1e-13}\) and \(g_\text{a\textgamma} = \SI{1e-12}{GeV^{-1}}\) for which axion-electron \(g_{ae}\) production is dominant.\cite{redondo2013solar, JvO_axionElectron}}
\end{figure}
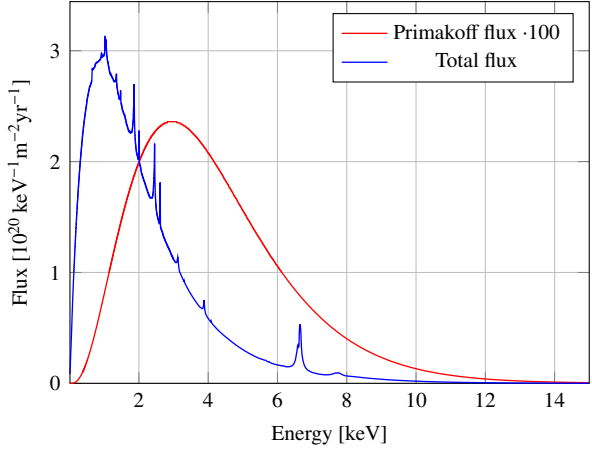

In order to search for solar axions, helioscopes like the
\textbf{C}ERN \textbf{A}xion \textbf{S}olar \textbf{T}elescope
(CAST)~\cite{ZIOUTAS1999480,PhysRevLett.94.121301} and its successor
the \textbf{I}nternational \textbf{AX}ion \textbf{O}bservatory
(IAXO)~\cite{Irastorza_2011,vogel2013iaxo,Armengaud_2014} are a very
promising approach. An axion helioscope mainly consists of a magnet
capable of following the Sun. The expected axion energy spectrum from the Sun for two
different production mechanisms is shown in figure~\ref{fig:axion_spectrum}:
\begin{itemize}
    \item axion production via their coupling to photons (Primakoff effect) peaks at energies between $\SIrange{0.5}{8}{keV}$ (red).
    \item axion production through the (model-dependent) coupling to electrons, peaks at energies between $\SIrange{0.5}{3}{keV}$ (blue)~\cite{redondo2013solar}.
\end{itemize}
The transverse magnetic field of the magnet is used to convert incoming axions to
photons of the same energy via the inverse Primakoff effect due to the
axion-photon coupling $g_\text{a\textgamma}$. To detect these photons, detectors sensitive in the very low energy X-ray regime are required. In addition, the expected values for $g_\text{a\textgamma}$ are very small, and thus the expected signal rates are very low. Therefore, a detector needs a very high detection efficiency at very low backgrounds. Here we report on a detector which satisfies these criteria and which was operated at the CAST experiment.

The CAST experiment consists of a decommissioned LHC prototype magnet,
providing two $\SI{9.26}{m}$ long bores with a field of up to
$\SI{9}{T}$. Two X-ray optics~\cite{kuster2007x, aznar2015micromegas}
are used on one end of the magnet bores in order to focus the X-rays
onto a small area where a detector can be mounted.

The detector is built as a pathfinder for the future IAXO experiment,
testing features for background reduction, signal enhancement, and
reliability. First, the detector shall be described in
section~\ref{sec:detector}.  Then, in section~\ref{sec:cdl}, the
calibration measurement using a variable X-ray generator will be
presented. After that, in section~\ref{sec:cast}, the data taking
campaign at the CAST experiment will be introduced, while at the end,
in section~\ref{sec:background}, the observed background rate of the detector
will be discussed. The influence of the different detector components is evaluated.

\section{Detector setup}
\label{sec:detector}
For the detection of solar axions via the
detection of low-energy X-ray photons, a specialized detector is
necessary. Therefore, a Micromegas-like detector, the GridPix-based
X-ray detector, was developed and operated successfully
\cite{krieger17_gridpix,krieger2018energy,krieger2018search,krieger_chameleon_jcap},
among others, at the CERN Axion Solar Telescope (CAST).

Starting from this, an upgraded detector with major improvements was developed. The detector consists of a Micromegas-like readout plane consisting of seven GridPixes (section~\ref{sec:detector:gridpix}), which is kept in a gas volume enclosed by a driftring with a field cage (section~\ref{sec:detector:driftring}). The detector is connected to the vacuum system of the X-ray optic via an ultra-thin X-ray entrance window (section~\ref{sec:detector:window}). If an X-ray enters the gas volume, it will produce a photoelectron, which, depending on its energy, will produce a certain amount of primary electrons. Due to an electrical field applied between the cathode with the entrance window and the
readout plane, the electrons will drift towards the readout through the gas, undergoing diffusion. At the readout plane, the electrons will then be amplified and measured by the GridPixes.  While keeping the readout principle, most improvements, such as a GridPix veto ring (section~\ref{sec:detector:septem}), two veto scintillators (section~\ref{sec:detector:scintis}) and a read out of the grid voltage via a Flash~ADC (FADC, section~\ref{sec:detector:fadc}) were introduced to yield an improvement of the background rejection.

\begin{figure}[htb]
  \centering
  \begin{tikzpicture}[font=\footnotesize]
    \node[anchor=south west,inner sep=0] (Bild) at (0,0) {\includegraphics[width=0.75\linewidth]{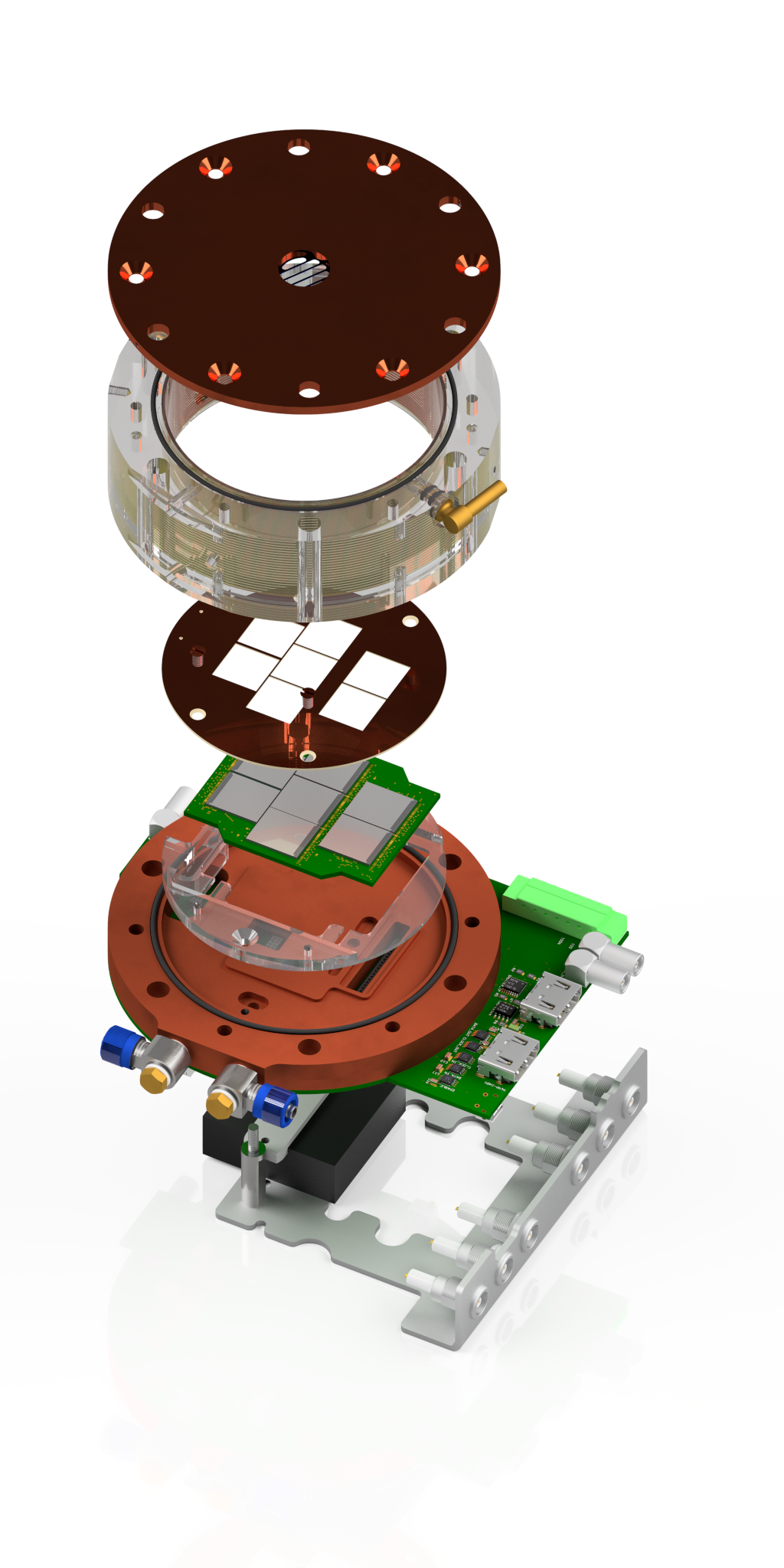}};
    \begin{scope}[x=(Bild.south east),y=(Bild.north west)]
      \node [right, align=left] (cat) at (0.9,0.825){Cathode with\\ entrance window};
      \draw [-stealth, thick, black](cat.west) -- (0.63,0.85);
      \node [right, align=left] (drift) at (0.9,0.7){Driftring with\\ field cage and\\ gas connectors};
      \draw [-stealth, thick, black](drift.west) -- (0.64,0.7);
      \node [right] (anode) at (0.9,0.6){Anode};
      \draw [-stealth, thick, black](anode.west) -- (0.57,0.57);
      \node [right] (sept) at (0.9,0.525){GridPix module};
      \draw [-stealth, thick, black](sept.west) -- (0.53,0.5);
      \node [right] (cool) at (0.9,0.45){Cooling plate};
      \draw [-stealth, thick, black](cool.west) -- (0.61,0.45);
      \node [right, align=left] (int) at (0.9,0.375){Readout PCB};
      \draw [-stealth, thick, black](int.west) -- (0.7,0.34);
      \node [right] (szint) at (0.9,0.275){Szintillator};
      \draw [-stealth, thick, black](szint.west) -- (0.46,0.25);
      \node [right, align=left] (hv) at (0.9,0.2){High voltage\\ connection};
      \draw [-stealth, thick, black](hv.west) -- (0.64,0.17);
    \end{scope}
  \end{tikzpicture}
  \caption{\label{fig:detector_explosion}Exploded view of the detector showing the individual components.}
\end{figure}
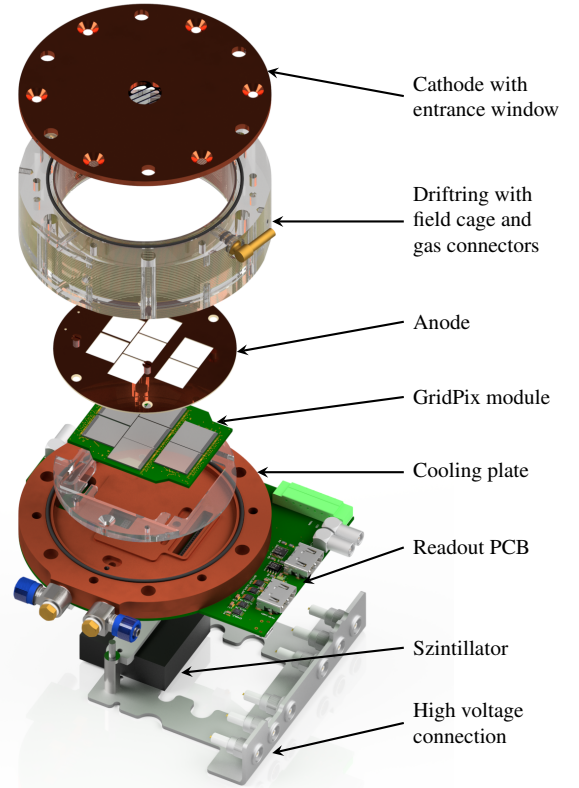

An overview of the detector is given in figure~\ref{fig:detector_explosion}. Most parts mentioned before can be seen, with the exception of the FADC and one veto scintillator. Further important parts are a dedicated cooling plate (section~\ref{sec:detector:cooling}) and (below that) the readout PCB (section~\ref{sec:detector:pcb}) which connects the detector to the readout electronics where the data are processed (section~\ref{sec:detector:signal_processing}). A detailed description can be found in~\cite{SchifferPhD}.

\subsection{GridPix}
\label{sec:detector:gridpix}
\begin{figure}[htbp]
  \centering
  \includegraphics[width=.9\linewidth]{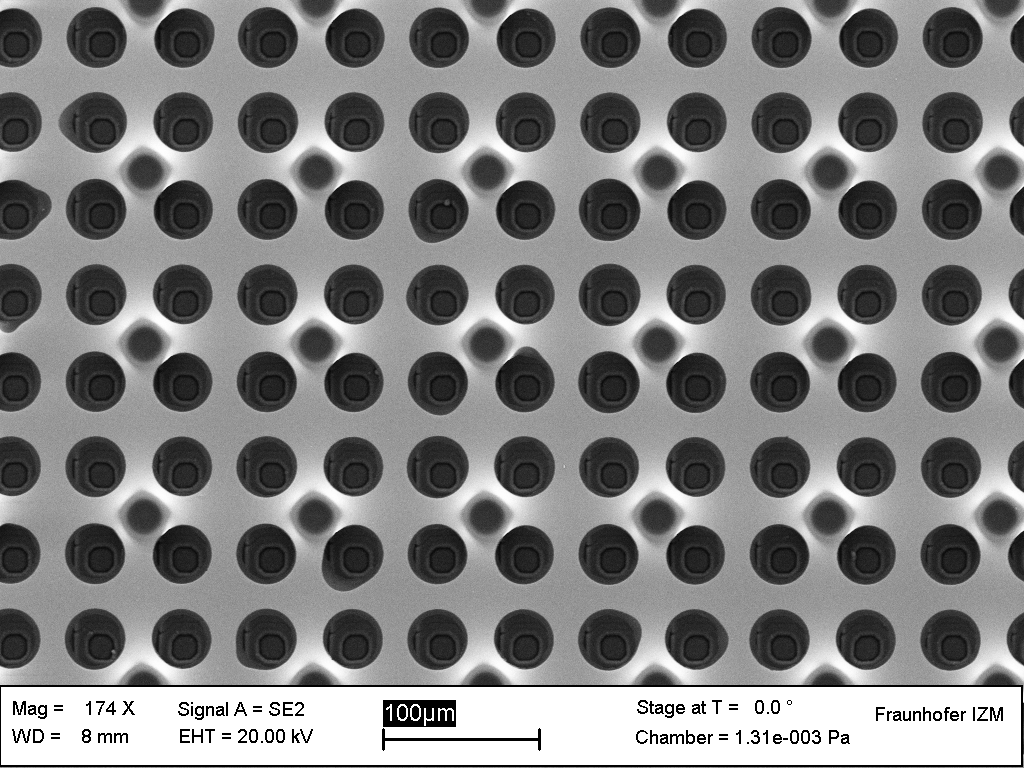}
  \caption{\label{fig:ingrid}SEM image of the grid above a Timepix ASIC. The grid and the pixels underneath the holes can be seen. Darker areas on the grid indicate the positions of the pillars holding the grid. Courtesy of~\cite{gridpix-izm-berlin}.}
\end{figure}

GridPix detectors are an evolution of the Micromegas technology, where a fine mesh (grid), used as an amplification stage, is produced via photolithographic post-processing on top of a pixelized readout
application-specific integrated circuit (ASIC), the Timepix
\cite{medipix,campbell2005detection,CHEFDEVILLE2006490,vanderGraaf:2007zz}. This
allows for small feature sizes, a perfect alignment of grid holes with pixels on the ASIC and therefore a very good spatial resolution. An SEM image is shown in figure~\ref{fig:ingrid}. The detection concept is similar to that of Micromegas: between the grid and the readout ASIC a high voltage is applied and the gap between the grid and the ASIC is filled with a gas. Incoming electrons are accelerated and due to the gas amplification are multiplied ($\mathcal{O}(10^3)$). This leads to a measurable charge signal at the input pads of the readout ASIC.  Due to diffusion and the good spatial resolution of the GridPix, typically each drifting electron is seen by a single pixel. If two or more drifting electrons reach the same pixel, the charge will be higher, which can be seen in the Time-over-Threshold value (ToT, a measure of charge) of the pixel. Therefore, there are two ways of determining the energy of the incident X-ray photon: counting the number of hit pixels or measuring the total charge. With this technique, photon energies down to the $\SI{200}{eV}$ range are measurable. This enables detection of solar axions in the regions where the flux of axions produced via the axion-electron coupling is dominating. (see figure~\ref{fig:axion_spectrum}).

\subsection{Driftring and field cage}
\label{sec:detector:driftring}
The driftring surrounds the gas volume. It supports
and isolates the cathode from the readout board and from the cooling
plate. At the same time it also holds the field cage, which is
necessary to form a homogeneous drift field over the full volume above
the seven GridPixes. The field cage consists of \num{29},
$\SI{0.7}{\milli\meter}$ wide, rings, separated by $\SI{0.3}{\milli\meter}$
spaces. The interconnection between the lines is realized by $\SI{10}{\mega\ohm}$ resistors representing a voltage divider chain. Thus, every line is on a different potential. The first and last ring are connected to high
voltage power supplies making the field cage voltages adjustable.

\subsection{Ultra-thin window}
\label{sec:detector:window}
\begin{figure}[htbp]
  \centering
  \includegraphics[width=0.5\linewidth]{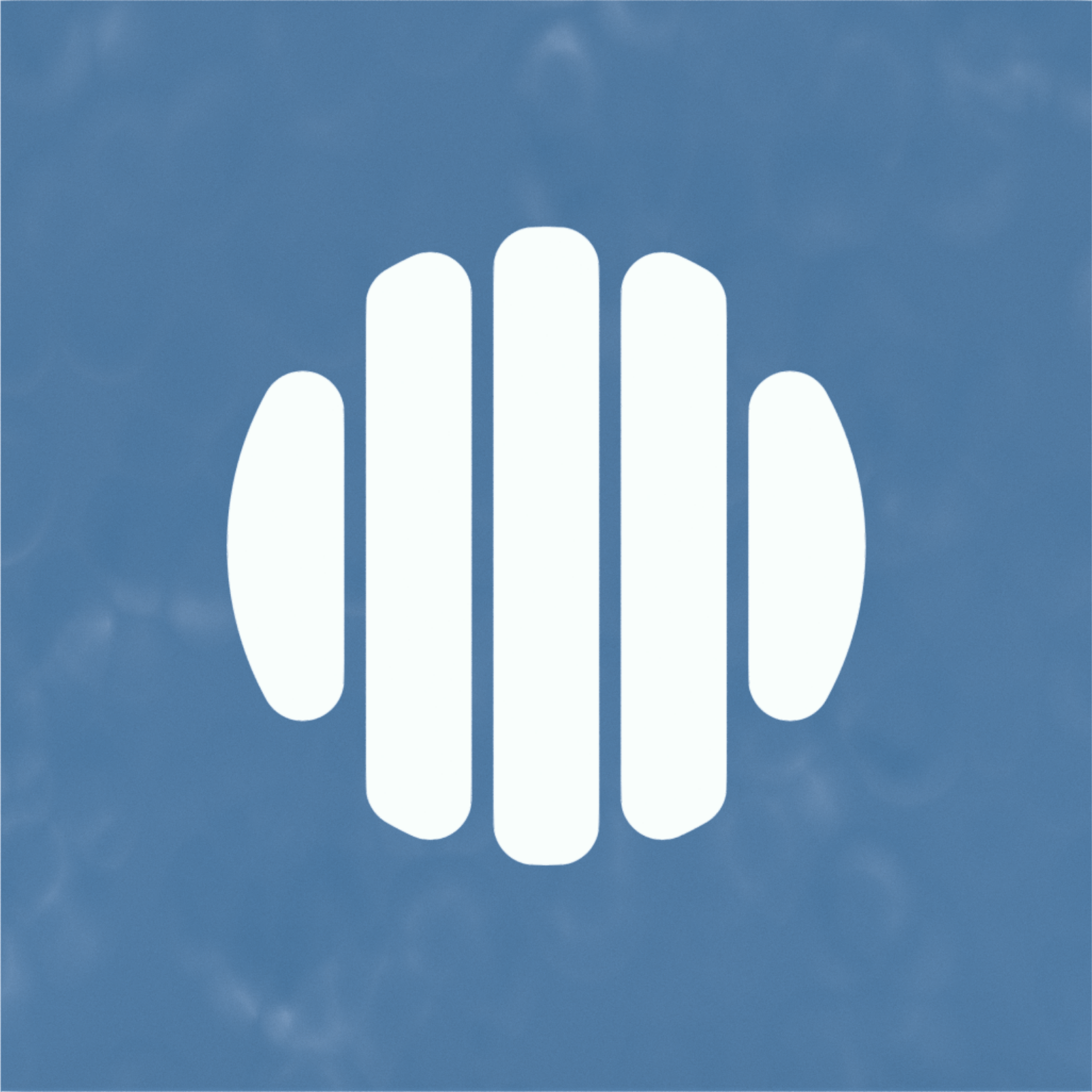}
  \caption{\label{fig:300nm_sin_norcada_window_layout}Layout of the
    \SI{300}{\nm} Norcada windows used at CAST in 2017/18. The
    strongback has a thickness of \SI{200}{\micro\meter} and thus is opaque
    to X-rays of desired energies.}
\end{figure}

The X-ray entrance window acts as the barrier between the vacuum
system and the gas filled volume of the detector. Thus, it needs to be
as transparent as possible for the incoming X-ray photons, while at
the same time being vacuum-tight and strong enough to withstand a
pressure difference of \SI{1050}{mbar}. To accomplish this goal we
developed, in cooperation with the company Norcada
\cite{gridpix-norcada}, ultra-thin silicon-nitride windows. The
windows have an open diameter of \SI{14}{\milli\meter} and a membrane
thickness of \SI{300}{\nano\meter} and a \SI{20}{nm} aluminium
coating. The membrane is supported by a strongback structure of four
ribs as shown in figure~\ref{fig:300nm_sin_norcada_window_layout},
which is taking up \SI{16.2}{\percent} of the open area. The windows
are pressure tested to survive at least six cycles up to a pressure
difference of \SI{1500}{mbar}, meaning they are very stable and do not
suffer significantly from fatigue stress. The measured helium leak
rate is lower than $\SI{3e-9}{mbar.l.s^{-1}}$. For details on the windows see~\cite{SchifferPhD}.

\begin{figure}[htb]
\centering
\begin{tikzpicture}[font=\footnotesize]
	\begin{axis}[
            axis y line*=right,
			axis x line=none,
            ylabel={Axion flux in a.u.},
            ytick scale label code/.code={},
			xmin=0.01, xmax=3.5,
			ymin = 0,
			scaled x ticks=false,
			width=0.95\linewidth,
			height=0.75\linewidth
			]
		\addplot[blue, line width=0.5pt] table[skip first n=1, x expr =(\thisrowno{0}/1000), y expr =(\thisrowno{9}/3.13), col sep=space] {axion_flux.csv};\label{plot_axionflux}
	\end{axis}
    \begin{axis}[
            grid = major,
            xlabel={Energy [keV]},
            ylabel=Transparency,
			xmin=0.01, xmax=3.5,
			ymin = 0, ymax = 1.1,
			scaled x ticks=false,
			width=0.95\linewidth,
			height=0.75\linewidth,
		      legend style={at={(0.85,0.03)},anchor=south east,cells={align=left}}
            ]

		\addplot[red, line width=0.5pt] table[skip first n=1, x expr =(\thisrowno{0}/1000), y expr = \thisrowno{1}*\thisrowno{6}, col sep=space] {SiN_window_trans.csv};
		\addlegendentry{$2\,\text{\textmu m}$ Mylar + $20\,\text{nm}$ Al}
		\addplot[green!60!black, line width=0.5pt] table[skip first n=1, x expr =(\thisrowno{0}/1000), y expr = \thisrowno{1}*\thisrowno{3}, col sep=space] {SiN_window_trans.csv};
		\addlegendentry{$300\,\text{nm}$ SiN + $20\,\text{nm}$ Al}
		\addlegendimage{/pgfplots/refstyle=plot_axionflux}\addlegendentry{Relativ axion flux}
    \end{axis}

\end{tikzpicture}
\caption{\label{fig:window_transmission} Comparison of two different
  window setups. A $\SI{2}{\micro m}$ thick Mylar window as it was
  used in the GridPix detector at CAST in 2014/15
  \cite{krieger2018search} and the \SI{300}{nm} SiN window setup used
  on this detector. As a reference the solar axion flux for an
  axion-electron coupling $g_{ae}$ dominated model is shown,
  highlighting the significant gain in transmission in the relevant
  energy ranges below
  $\SI{3}{keV}$~\cite{Schmidt_xrayAttenuation_2022,
    JvO_axionElectron}.}
\end{figure}
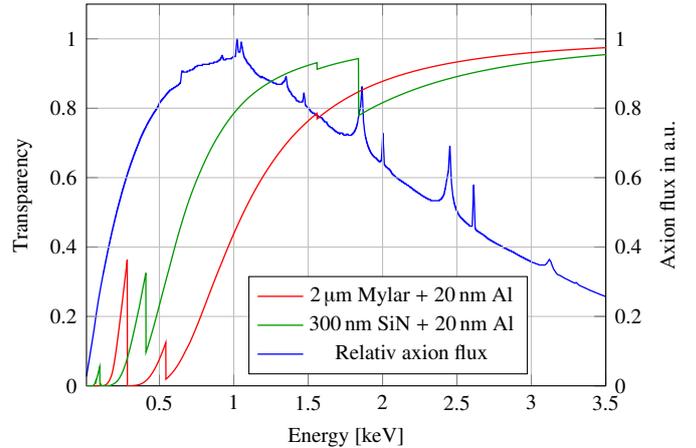

The main purpose of thin windows is the increase in X-ray
transmission, in particular at low
energies. Figure~\ref{fig:window_transmission} shows the transmission of
different window setups \footnote{The transmissions were calculated
with \texttt{xrayAttenuation} \cite{Schmidt_xrayAttenuation_2022}
using data from NIST \cite{hubbell1996nist} and the Lawrence
Berkeley National Laboratory \cite{henke1993x}.}. Note in particular
the improved transmission of the silicon-nitride (SiN) setup compared to
the Mylar setup with respect to the axion-electron flux.

\subsection{Seven GridPixes}
\label{sec:detector:septem}
The seven-GridPix module consists of one central chip for the
measurement of the converted axions surrounded by six veto GridPixes to
reduce the background from cosmic particles further. This is
especially helpful for the outer parts of the central chip, where a
traversing muon can easily mimic a photon-like spherical cluster. The outer six
GridPixes are arranged around the central chip, leaving a
gap to allow wire bonding of the middle row. The layout is shown in
figure~\ref{fig:detector_explosion}.

\subsection{FADC trigger}
\label{sec:detector:fadc}
Signals induced by back-drifting ions on the grid of the central GridPix are
decoupled from the high voltage connection using a capacitor. These
voltage pulses are amplified and sent to a
'Flash ADC' (FADC).

The FADC samples the signal from the grid with a resolution of
\SI{1}{\nano\second}. These signals can be used to differentiate
between real photon events and muons entering the detector
perpendicularly to the chip, yielding photon-like events. Muon-induced
ionization of the gas happens along a track (the electrons arrive over
time at the grid), while in the case of the photon the ionization
originates at a single point (electrons arrive roughly at the same
time). This leads to a difference in the time structure of the signal.

The second use case of the FADC is sending out a trigger signal as
soon as it sees a signal. This trigger is then used to readout the
current frame of the detector, thus reducing the likelihood of frames
with more than one primary physical interaction. This allows for
longer shutter times increasing the duty cycle of the detector
drastically.

Unfortunately, due to the decoupled signals from the grid being very
small, the high-gain amplification of the FADC trigger system can easily
pick up noise. This restricts the threshold of the FADC to
$\sim\SI{2}{keV}$ to avoid noise triggering a readout. As such,
below the FADC trigger energy, it cannot act as a GridPix readout trigger. However also those signal will be read out after the regular shutter time.

\subsection{Cooling}
\label{sec:detector:cooling}
The dissipated heat per GridPix can reach $\sim\SI{2}{\watt}$ in
typical applications, yielding \SI{14}{\watt} of possible heat
production in total for seven GridPixes. Due to the very small volume
of gas and most parts being covered in plastic, this heat is not
radiated out of the detector and instead leads to a significant heat
build-up. This has a number of consequences.

First, the gas amplification $G$ rises with temperature
$G\propto \text{e}^T$, meaning a stable operation is not possible
under varying temperatures.  Second, after exceeding a threshold of
approximately \SI{95}{\celsius} sparks start to occur. These sparks
lead to dead time. However, already at much lower temperatures
($\sim\SI{65}{\celsius}$) the Timepix ASICs start to be unstable due
to drifts in the threshold DAC and produce significant noise. At even
higher temperatures the ASIC can be destroyed.

Therefore, a cooling device was necessary for the detector. The
possible thickness of the cooling device is limited by the small
amount of space in the lead shielding at the CAST setup. Further, the
cooling needs to be standalone since it needs to be set up on a
moving magnet. A closed-loop water cooling device was built (see figure~\ref{fig:detector_explosion})
consisting of a copper cooling plate to which the
GridPixes are thermally coupled. This is connected to a water
reservoir and an air-cooled radiator. The water is circulated with a
small pump sitting behind the reservoir. Close to the GridPixes, a
temperature sensor (\texttt{PT 1000}) is mounted to measure the temperature
close to the chip, used in an interlock system to avoid destruction by
overheating.

\subsection{Readout PCB}
\label{sec:detector:pcb}
The readout PCB provides the high voltage lines to the grids of the
GridPixes and the back mounted scintillator. In addition, it powers
the GridPixes and the temperature readout.

Further, it connects the seven GridPixes and the back-mounted veto
scintillator to a Field Programmable Gate Array (FPGA) board (Virtex-6
FPGA ML605 Evaluation Kit
\footnote{\url{https://www.xilinx.com/products/boards-and-kits/ek-v6-ml605-g.html}})
via a custom made adapter card. On the FPGA, all the data is processed
and then sent to the PC.

The chip data is sent from the readout PCB to the FPGA board via two
HDMI cables. One line pair is used for an \texttt{I${}^\mathtt{2}$C} signal, which is split
up to set the readout mode of the GridPixes. Additionally, a coaxial
line is used to connect the back-mounted veto scintillator to the
adapter card. The adapter card has additional coaxial connections to
feed the second veto scintillator and the FADC trigger signal
into the FPGA.

The temperature readout uses a USB interface connected to the
PC. Two temperature sensors are used, one on the readout PCB as reference and
one close to the GridPixes inside the detector.

\subsection{Scintillator veto}
\label{sec:detector:scintis}
The scintillator veto consists of two scintillators: a small
scintillator (SiPM) of $\SI[parse-numbers=false]{4 \times 4 \times 1}{cm^3}$
is sitting behind the central GridPix. This one is
used to reject muons approaching the chip from the front. The second
scintillator is \(\SI[parse-numbers=false]{80 \times 40 \times 5}{cm^3}\) and set
up above the detector, the lead shielding, and the first part of the
beamline in order to suppress events induced by muons hitting from
above.

Cosmic Muons can cause X-ray fluorescence from the detector/beamline materials
introducing signal-like events in the detector.  Both scintillators
are used in a way that they tag a frame if one of them has triggered
within \(\SI{100}{\micro s}\) before the frame was stopped by the FADC
trigger, meaning that an event was recorded. All such frames can then
optionally be rejected offline, where typically a cut-off of
\(\SI{3.75}{ \micro s}\) is used.

\begin{figure}[htbp]
  \begin{subfigure}{\linewidth}
  \centering
    \includegraphics[width=0.9\linewidth]{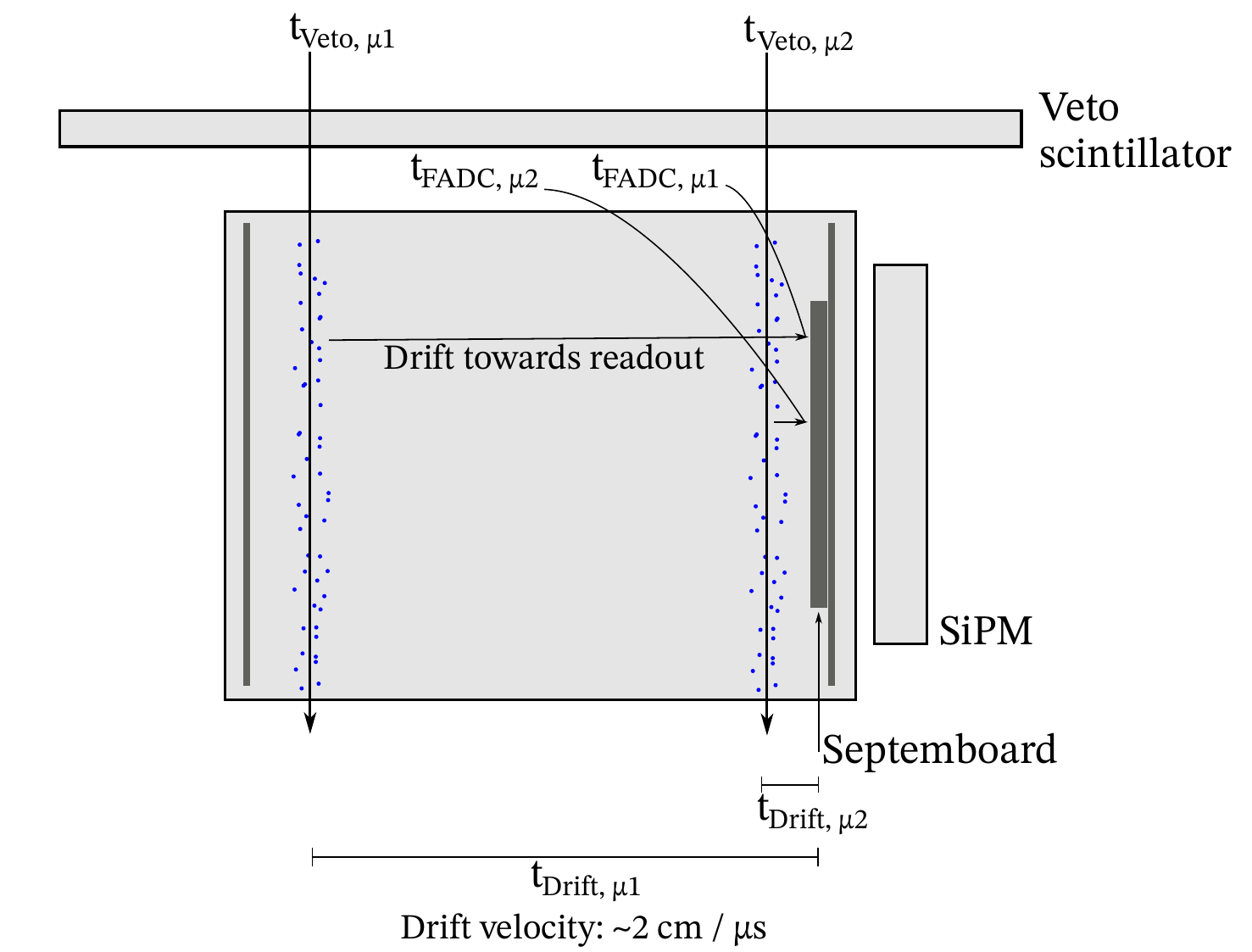}
    \caption{Veto paddle}
    \label{fig:detector:fadc_veto_paddle_expl}
  \end{subfigure}\\
  \begin{subfigure}{\linewidth}
  \centering
    \includegraphics[width=0.9\linewidth]{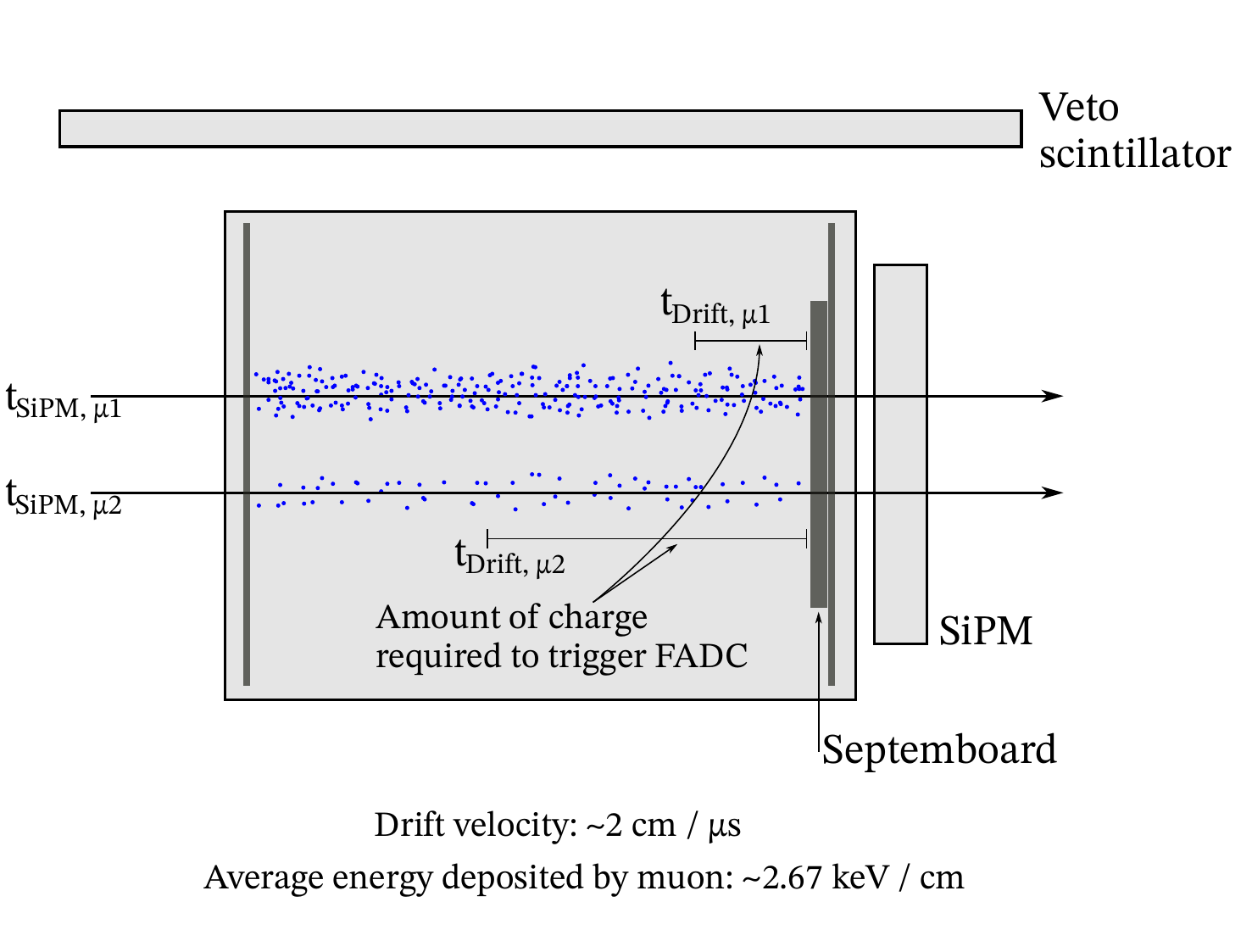}
    \caption{SiPM}
    \label{fig:detector:fadc_sipm_expl}
  \end{subfigure}%
  \caption{(\subref{fig:detector:fadc_veto_paddle_expl}): Schematic of
    expected signals for different muons from the zenith passing
    through scintillator paddle and detector. $t_{\text{Veto}}$ marks
    beginning of a counter.(\subref{fig:detector:fadc_sipm_expl}): Ionization of a muon
    is a statistical process. Depending on the density of the charge
    cloud for muons orthogonal to the readout plane, the time to accumulate
    enough charge to trigger the FADC differs.}
  \label{fig:detector:fadc_scintillators_explanation}
\end{figure}

A schematic showcasing of the ideas behind each of the two scintillators
is shown in figure~\ref{fig:detector:fadc_scintillators_explanation}. In figure~\ref{fig:detector:fadc_veto_paddle_expl} additionally the cutoff value of \(\SI{3.75}{ \micro s}\) for the delay between the trigger of the FADC and the Scintillator signal due to the drift velocity of the electrons is explained.

\subsection{Signal and veto processing on FPGA}
\label{sec:detector:signal_processing}
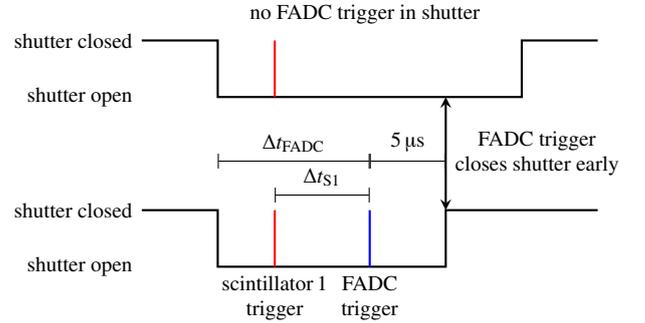
\begin{figure}[htb]
  \centering
  \begin{tikzpicture}[font=\footnotesize]
      \node [right] at (1.3,0.35){no FADC trigger in shutter};
      \node [left] (shutter_closed1) at (0,0){shutter closed};
      \node [left] (shutter_open1) at (0,-0.75){shutter open};
      \node [left] (shutter_closed1) at (0,-2.25){shutter closed};
      \node [left] (shutter_open1) at (0,-3){shutter open};
      \draw[thick, black](0,0) -- (1,0)--(1,-0.75)--(5,-0.75)--(5,0)--(6,0);
      \draw[thick, black](0,-2.25) -- (1,-2.25)--(1,-3)--(4,-3)--(4,-2.25)--(6,-2.25);
      \draw[stealth-stealth, thick, black](4,-2.25) -- (4,-0.75);
      \draw[thick, red](1.75,-3) -- (1.75,-2.25);
      \draw[thick, red](1.75,0) -- (1.75,-0.75);
      \draw[thick, blue](3,-3) -- (3,-2.25);
      \draw[|-|, black](1.75,-2.05) -- (3,-2.05);
      \draw[|-|, black](1,-1.6) -- (3,-1.6);
      \draw[|-|, black](4,-1.6) -- (3,-1.6);
      \node[below, align=center] at (3,-3) {FADC \\ trigger};
      \node[below, align=center] at (1.75,-3) {scintillator\,1 \\ trigger};
      \node[right, align=center] at (4,-1.5) {FADC trigger \\ closes shutter early};
      \node[above] at (3.5,-1.6) {{$\SI{5}{\micro s}$}};
      \node[above] at (2,-1.6) {{$\Delta t_\text{FADC}$}};
      \node[above] at (2.375,-2.05) {{$\Delta t_\text{S1}$}};
  \end{tikzpicture}
  \caption{\label{fig:detector:scintillator_fadc_shutter_close}Schematic showing how the FADC and scintillators are used together to tag possible coincidence events and close the shutter early to reduce the likelihood of multi-hit events. If the scintillator triggers when the shutter is open (red line), a clock starts counting up to 4000 clock cycles. On every new trigger this clock is reset. If the FADC triggers (blue line), the scintillator clock values (e.g. $\Delta t_\text{S1}$) and the time the shutter was open ($\Delta t_\text{FADC}$) are read out and can be used to correlate events in the scintillator with FADC and GridPix information. Further, the FADC trigger is used to close the Timepix shutter \(\SI{5}{\micro s}\) after the trigger.}
\end{figure}

The incoming and
outgoing signals for the communication between the computer and the
GridPixes are processed on the FPGA: for communication towards the
GridPixes, the FPGA generates the necessary control signals and sends
those via HDMI cables to the GridPixes. Parts of these control signals
are sent via \texttt{I${}^\mathtt{2}$C} to reduce the number of data lines. The
\texttt{I${}^\mathtt{2}$C} signals are also generated on the FPGA and then processed
on the readout board by an \texttt{I${}^\mathtt{2}$C} expander. After a predefined shutter time, readout is performed as shown in the top part of figure~\ref{fig:detector:scintillator_fadc_shutter_close}. The data coming from
the GridPixes is daisy-chained, therefore only one differential line is
needed for this. The received data is then zero-suppressed on the FPGA
for all GridPixes, allowing a maximum pixel count of \num{4096} per
GridPix. If there is more data on one GridPix it is cut off. The
zero-suppressed data is then sent to the computer to be further
analyzed.

Additionally a trigger can be used to read out the GridPixes
earlier (figure~\ref{fig:detector:scintillator_fadc_shutter_close} bottom part). The FADC trigger is used for this and will stop any frame
\(\SI{5}{\micro s}\) after the FADC is triggered (blue line). The
\(\SI{5}{\micro s}\) are necessary to enable the full event to reach
the readout and to allow for a proper ToT measurement. An additional clock
signal $\Delta t_\text{FADC}$ gives information about the time during which the
shutter was open, as shown in
figure~\ref{fig:detector:scintillator_fadc_shutter_close}. While the
shutter is open, two additional clocks ($\Delta t_\text{S1}$ and
$\Delta t_\text{S2}$) can be started by the veto scintillators
(e.g. red line for $\Delta t_\text{S1}$). These clocks run for up to
\SI{100}{\micro\second} with a resolution of \SI{25}{\nano\second}. If
more time has passed before the FADC trigger, a special value is sent,
indicating that there was a veto trigger signal in the frame, but more
than \SI{100}{\micro\second} ago. Each signal from the veto
scintillator will start the clock again from zero so only the most recent
trigger will be saved. With each FADC trigger those three clock values
will be, in addition to the frame data, sent to the computer. If the
FADC does not trigger, no timestamps are sent to the computer.

\section{CAST detector Lab}
\label{sec:cdl}
\begin{figure}[htbp]
  \begin{subfigure}{\linewidth}
  \centering
    \includegraphics[width=0.8\linewidth]{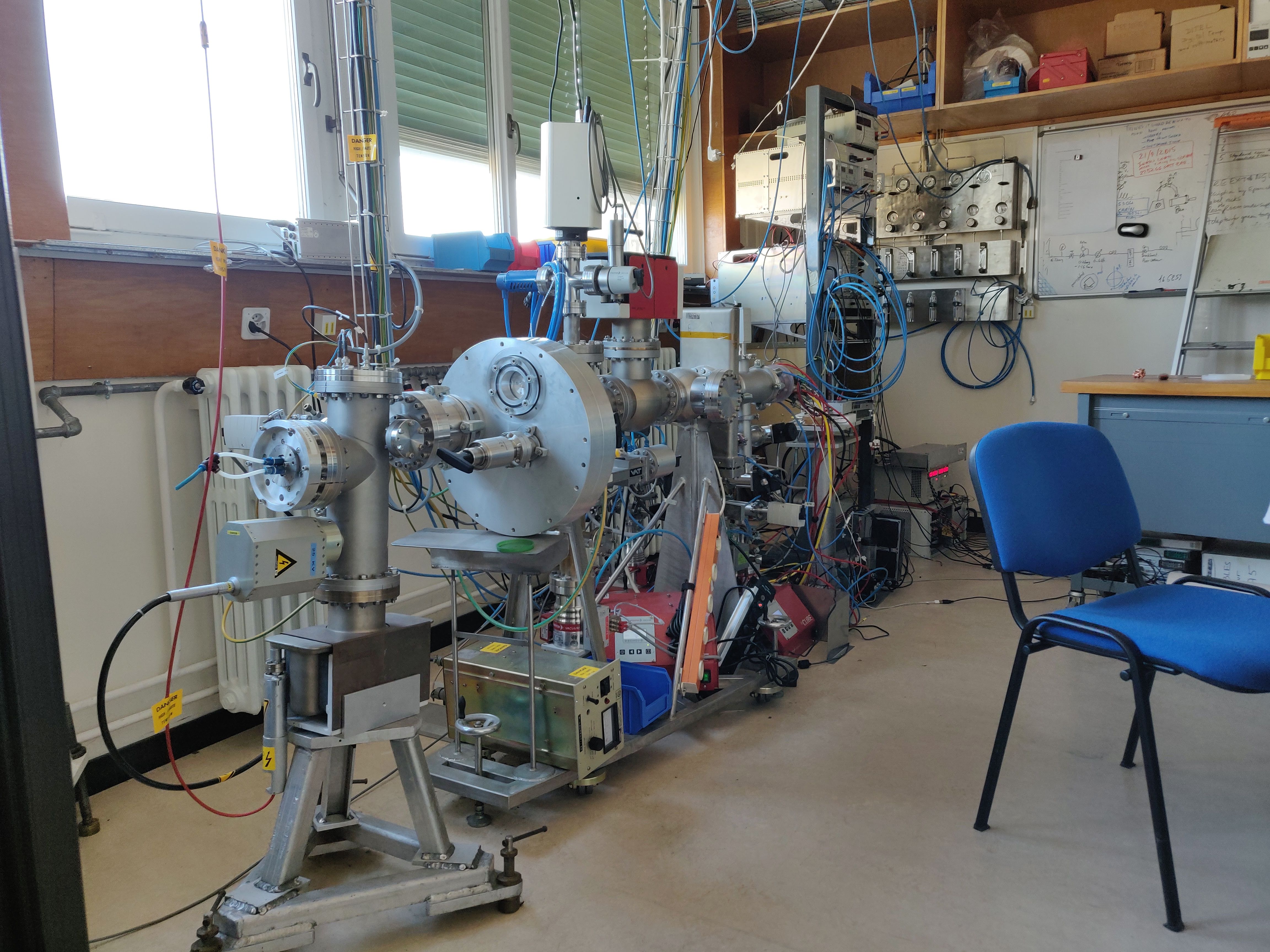}
    \caption{Full setup}
    \label{fig:cdl:cdl_setup}
  \end{subfigure}\\
  \begin{subfigure}{\linewidth}
  \centering
    \includegraphics[width=0.8\linewidth]{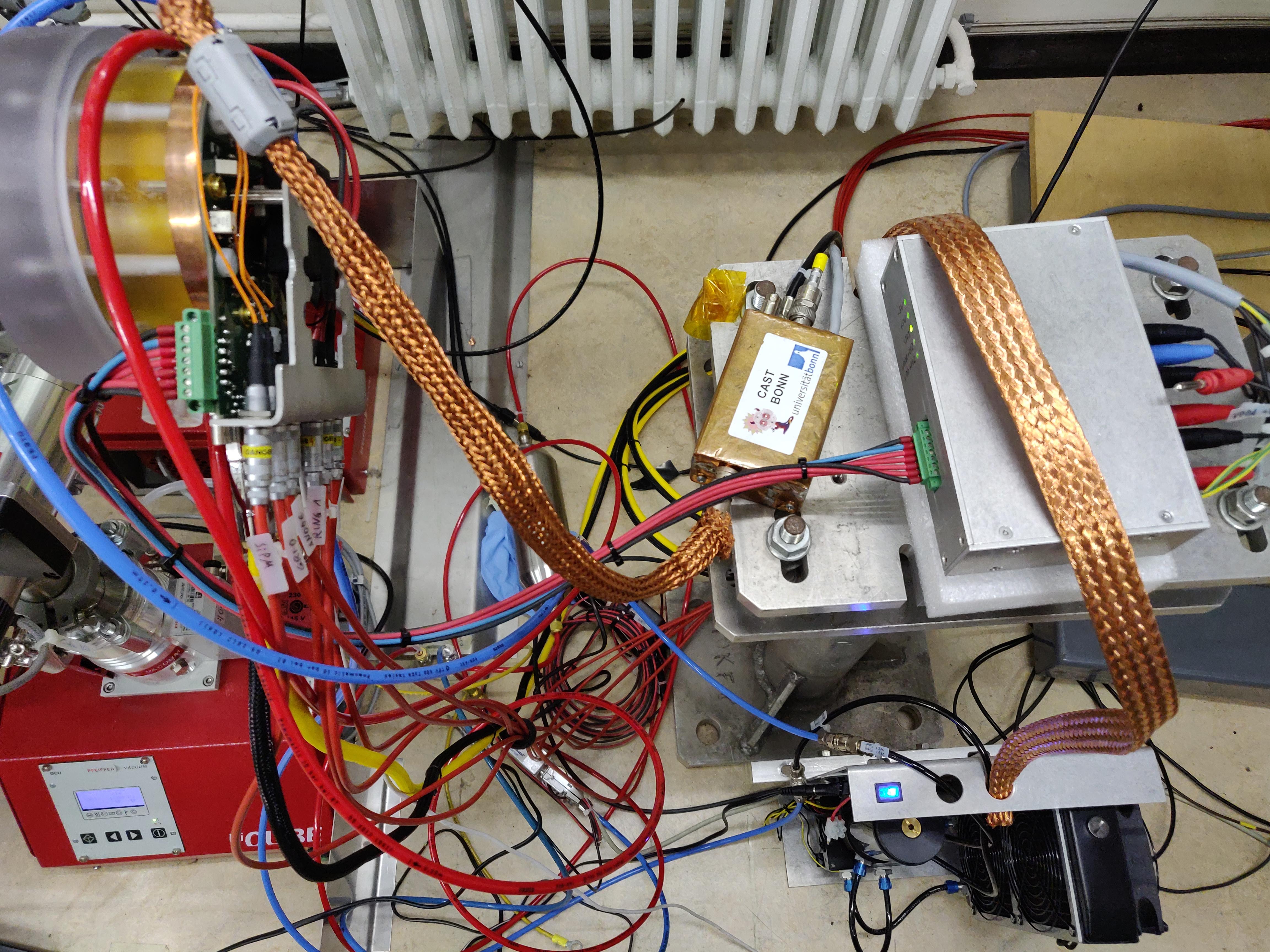}
    \caption{Detector setup}
    \label{fig:cdl:detector_setup}
  \end{subfigure}%
  \caption{(\subref{fig:cdl:cdl_setup}) shows the full vacuum test stand containing the X-ray tube with
    the seven-GridPix detector installed at the rear, visible by the red HV and yellow HDMI cables. (\subref{fig:cdl:detector_setup}) is a view of the detector setup from above. On the left hand side
    the detector mounted to the vacuum setup is shown. The water cooling is seen in the bottom right, connected via
    the blue tubes. The gas supply is in red tubing and the power supply is visible on the right above the
    water cooling (with a green Phoenix connector). The copper shielded cable is a LEMO cable for the FADC
    signal going to the pre-amplifier (with the University of Bonn sticker).}
  \label{fig:cdl:cdl}
\end{figure}

The detector described here has been tested and calibrated with X-rays of several energies at the \textbf{C}AST \textbf{D}etector \textbf{L}ab (CDL) at CERN. The CDL has an X-ray
source with variable targets followed by a filter wheel. By combining
a certain high voltage with a target and filter, specific X-ray
energies can be transported to the device under test. Since the system
is set up for low-energy X-rays, the system is completely under vacuum
when operated. To reduce the pressure difference in front of the window slowly,
additional valves, including a needle valve, have been introduced
to the system. Figure~\ref{fig:cdl:cdl_setup} shows an image of the
setup, while figure~\ref{fig:cdl:detector_setup} shows the detector,
power supply and water cooling installed to the system.

The X-ray energies and target / filter combinations used for this
detector are shown in table~\ref{tab:cdl_lines}. Over the range of
\(\SIrange{0.277}{8.04}{keV}\), eight X-ray energies were used for
calibration, so the range of the expected axion signals is covered
very well.

For each setting $\sim 35000$ events were recorded, of which $\sim 10000$
pass cuts (e.g remove double hits via their eccentricity) in order to obtain a clean X-ray
dataset of the targeted energy. These datasets are used to define the
properties of the X-ray clusters, at the energy of the flourescence
line used, based on a set of geometric variables. For energies between
two lines a linear interpolation of the distributions is
performed. Outside the range of available X-ray energies, the lowest / highest distribution is used independent of energy.

\begin{figure}[htb]
\centering
\begin{tikzpicture}[font=\footnotesize]
    \begin{axis}[
                grid = major,
                xlabel={Charge $[10^5\,\text{e}^-]$},
                ylabel=Counts,
				xmin=0, xmax=14,
			    ymin=0, ymax=1000,
				scaled y ticks=false,
				scaled x ticks=false,
				width=0.95\linewidth,
				height=0.6\linewidth,
				legend style={at={(0.97,0.97)},anchor=north east,cells={align=left}}
                ]
    \addplot[green!60!black,ybar interval, draw=none, fill, opacity=0.4,area legend] table[x expr=\thisrowno{0}/100000+0.05, y index=1, col sep=space] {Ti_Ti_hist.txt};
    \addlegendentry{Raw data}
	\addplot[purple!80!black, ybar interval, draw=none, fill, opacity=0.4,area legend] table[x expr=\thisrowno{0}/100000+0.05, y index=2, col sep=space] {Ti_Ti_hist.txt};
    \addlegendentry{Cut data}
    \addplot[purple!60!black, line width=1pt] table[skip first n=1, x expr = \thisrowno{1}/100000, y expr = \thisrowno{2}, col sep=comma] {Ti-Ti_fit.csv};
    \addlegendentry{Fit}
    \addplot[ samples=5, smooth, black, name path = cut1] coordinates {(3.14253,0)(3.14253,400)};
	\addplot[ samples=5, smooth, black, name path = cut2] coordinates {(6.09747,0)(6.09747,400)};
    \addplot[ samples=5, smooth, draw=none, name path = cut1err1] coordinates {(3.14253-3*0.02441,0)(3.14253-3*0.02441,400)};
    \addplot[ samples=5, smooth, draw=none, name path = cut1err2] coordinates {(3.14253+3*0.02441,0)(3.14253+3*0.02441,400)};
	\addplot[ samples=5, smooth, draw=none, name path = cut2err1] coordinates {(6.09747-3*0.02441,0)(6.09747-3*0.02441,400)};
    \addplot[ samples=5, smooth, draw=none, name path = cut2err2] coordinates {(6.09747+3*0.02441,0)(6.09747+3*0.02441,400)};
	\addplot [fill=black, opacity=0.4]fill between[of=cut1err1 and cut1err2];
    \addplot [fill=black, opacity=0.4]fill between[of=cut2err1 and cut2err2];
	\end{axis}
\end{tikzpicture}
\caption{\label{fig:cdl:ti_ti_charge_spectrum_run_326}Charge spectrum
  of the \(\text{Ti}-\text{Ti}\) spectrum at \(\SI{9}{kV}\). The green
  histogram shows the raw data and the purple histogram indicating the
  data left after the cleaning cuts are applied. The purple line
  indicates the result of the fit as described in
  table~\ref{tab:cdl:fit_func_charge}. The black lines represent the
  \(3\sigma \) region around the main fluorescence line (with grey
  error bands), which is later used to extract those clusters likely
  from the fluorescence line and therefore known energy.}
\end{figure}
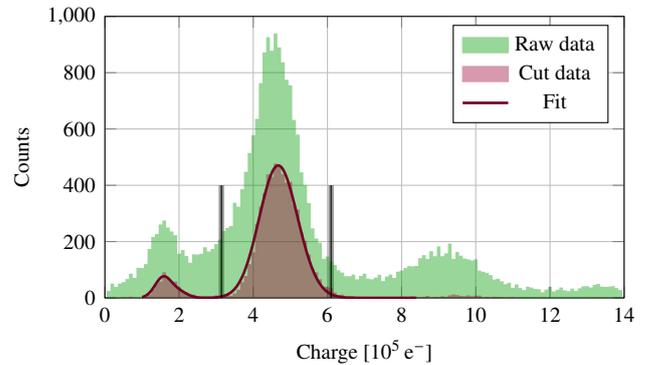

Each obtained spectrum is fitted with a sum of normal distributions
(one for each fluorescence line present for each target / filter
combination) to obtain the energy resolution and map the event charge and number of hit pixels values to energies. The functions fitted to the charge
spectra are shown in table~\ref{tab:cdl:fit_func_charge}. The event charge spectrum is used since it has a slightly better discrimination power, especially at small cluster sizes where double hits are possible. An example spectrum
for the \(\SI{4.51}{keV}\) Ti fluorescence line is shown in
figure~\ref{fig:cdl:ti_ti_charge_spectrum_run_326}. The green histogram shows
the raw events and the purple those passing the previously mentioned
filter cuts. The fit is shown in purple. Only X-rays
within the vertical, shaded bars around the main fluorescence line are
used for further definition of the geometry of X-ray clusters.
The low statistics of the calibration datasets comes from a limitation of the detector
design: the readout of all seven GridPixes is daisy-chained, which leads to significant
readout times \(\mathcal{O}(> \SI{200}{ms})\). When taking data with any
kind of active source, dead time due to readout dominates over X-ray
sensitive time. However, this does not affect the data taking at the experiment where event rates are very low.

\begin{table}[htb]
  \centering
  \begin{tabular}{llllrl}
    \hline
    Set & Target    & Filter    & HV       & Fluorescence & Energy \\
        &           &           &[\si{kV}]& &[\si{keV}]\\
    \hline
    1   & Cu & Ni    & 15 & Cu K\textalpha & 8.04 \\
    2   & Mn & Cr    & 12 & Mn K\textalpha & 5.89 \\
    3   & Ti & Ti    &  9 & Ti K\textalpha & 4.51 \\
    4   & Ag & Ag    &  6 & Ag L\textalpha & 2.98 \\
    5   & Al & Al    &  4 & Al K\textalpha & 1.49 \\
    6   & Cu & EPIC  &  2 & Cu L\textalpha & 0.930 \\
    7   & Cu & EPIC  &  0.9 & O K\textalpha  & 0.525 \\
    8   & C  & EPIC  &  0.6 & C K\textalpha  & 0.277 \\
    \hline
  \end{tabular}
  \caption{\label{tab:cdl_lines}Table of all target / filter combinations used with the described detector,
    the high voltages applied to the X-ray tube and the corresponding
    fluorescence lines with their energy. See \protect\footnotemark for
    information about the EPIC filter.}
\end{table}

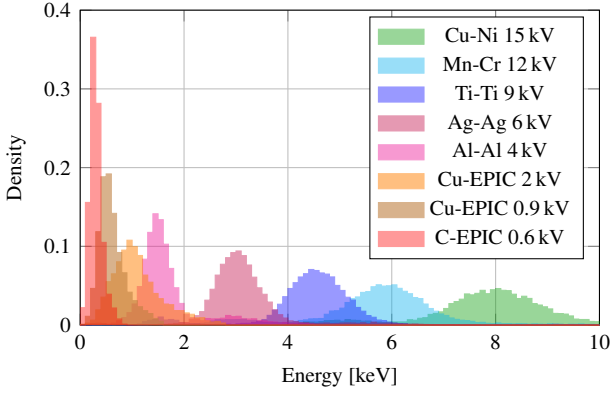
\begin{figure}[htb]
\centering
\begin{tikzpicture}[font=\footnotesize]
    \begin{axis}[
                grid = major,
                xlabel={Energy [keV]},
                ylabel=Density,
				xmin=0, xmax=10,
			    ymin=0, ymax=0.4,
				scaled y ticks=false,
				scaled x ticks=false,
				width=0.95\linewidth,
				height=0.65\linewidth,
				legend style={at={(0.97,0.98)},anchor=north east,cells={align=left}}
                ]
    \addplot[green!60!black,ybar interval,const plot, fill, opacity=0.4,area legend] table[skip first n=1, x index=0, y index=1, col sep=space] {CDL_hist.csv};
    \addlegendentry{Cu-Ni 15\,kV}
	\addplot[cyan, ybar interval,const plot, fill, opacity=0.4,area legend] table[skip first n=1, x index=0, y index=2, col sep=space] {CDL_hist.csv};
    \addlegendentry{Mn-Cr 12\,kV}
    \addplot[blue, ybar interval,const plot, fill, opacity=0.4,area legend] table[skip first n=1, x index=0, y index=3, col sep=space] {CDL_hist.csv};
    \addlegendentry{Ti-Ti 9\,kV}
    \addplot[purple, ybar interval,const plot, fill, opacity=0.4,area legend] table[skip first n=1, x index=0, y index=4, col sep=space] {CDL_hist.csv};
    \addlegendentry{Ag-Ag 6\,kV}
    \addplot[magenta, ybar interval,const plot, fill, opacity=0.4,area legend] table[skip first n=1, x index=0, y index=5, col sep=space] {CDL_hist.csv};
    \addlegendentry{Al-Al 4\,kV}
    \addplot[orange, ybar interval,const plot, fill, opacity=0.5,area legend] table[skip first n=1, x index=0, y index=6, col sep=space] {CDL_hist.csv};
    \addlegendentry{Cu-EPIC 2\,kV}
    \addplot[brown, ybar interval,const plot, fill, opacity=0.6,area legend] table[skip first n=1, x index=0, y index=7, col sep=space] {CDL_hist.csv};
    \addlegendentry{Cu-EPIC 0.9\,kV}
    \addplot[red, ybar interval,const plot, fill, opacity=0.4,area legend] table[skip first n=1, x index=0, y index=8, col sep=space] {CDL_hist.csv};
    \addlegendentry{C-EPIC 0.6\,kV}
	\end{axis}
\end{tikzpicture}
\caption{\label{fig:cdl:calibrated_energy_histos}Normalized histograms
  of all CDL data after applying basic cuts and calibrating the data
  in energy using the charge of the main fitted line and its known
  energy as a baseline. Some targets show a wider energy distribution due to
  varying detector conditions during the measurement.}
\end{figure}

\begin{table}[htb]
\centering
\begin{tabular}{ll}
\hline
Set & Fit functions\\[0pt]
\hline
1    & $G^{\text{Cu}}_{K\alpha} + G^{\text{Cu}, \text{esc}}_{K\alpha}$\\[0pt]
2    & $G^{\text{Mn}}_{K\alpha} + G^{\text{Mn}, \text{esc}}_{K\alpha}$\\[0pt]
3     & $G^{\text{Ti}}_{K\alpha} + G^{\text{Ti}, \text{esc}}_{K\alpha} + G^{\text{Ti}}_{K\beta}\left( \mu^{\text{Ti}}_{K\alpha}\cdot\left(\frac{4.932}{4.511}\right), \sigma^{\text{Ti}}_{K\alpha} \right)$ \\
        &              $+ G^{\text{Ti}}_{K\beta}\left( \mu^{\text{Ti}}_{K\alpha}\cdot\left(\frac{4.932}{4.511}\right), \sigma^{\text{Ti}}_{K\alpha} \right)$\\
        &               $+ G^{\text{Ti}, \text{esc}}_{K\beta}\left( \mu^{\text{Ti}}_{K\alpha}\cdot\left(\frac{1.959}{4.511}\right), \sigma^{\text{Ti}, \text{esc}}_{K\alpha} \right)$\\[0pt]
4    & $G^{\text{Ag}}_{L\alpha} + G^{\text{Ag}}_{L\beta}\left( N^{\text{Ag}}_{L\alpha}\cdot 0.56, \mu^{\text{Ag}}_{L\alpha}\cdot \left(\frac{3.151}{2.984}\right), \sigma^{\text{Ag}}_{L\alpha} \right)$\\[0pt]
5     & $G^{\text{Al}}_{K\alpha}$\\[0pt]
6     & $G^{\text{Cu}}_{L\alpha} + G^{\text{Cu}}_{L\beta}\left( N^{\text{Cu}}_{L\alpha}\cdot\left(\frac{0.65}{1.11}\right), \mu^{\text{Cu}}_{L\alpha}\cdot\left(\frac{0.9498}{0.9297}\right), \sigma^{\text{Cu}}_{L\alpha} \right) $\\
        &               $+ G^{\text{O}}_{K\alpha}\left( \frac{N^{\text{Cu}}_{L\alpha}}{3.5}, \mu^{\text{Cu}}_{L\alpha}\cdot\left(\frac{0.5249}{0.9297}\right), \frac{\sigma^{\text{Cu}}_{L\alpha}}{2.0} \right) + G_{\text{unknown}}$\\[0pt]
7   & $G^{\text{O}}_{K\alpha} + G^{\text{C}}_{K\alpha}\left( \frac{N^{\text{O}}_{K\alpha}}{10.0}, \mu^{\text{O}}_{K\alpha}\cdot\left(\frac{277.0}{524.9}\right), \sigma^{\text{O}}_{K\alpha} \right) + G_{\text{unknown}}$\\[0pt]
8   & $G^{\text{C}}_{K\alpha} + G^{\text{O}}_{K\alpha}\left( \mu^{\text{C}}_{K\alpha}\cdot\left(\frac{0.525}{0.277}\right), \sigma^{\text{C}}_{K\alpha} \right)$\\[0pt]
\hline
\end{tabular}
\caption{\label{tab:cdl:fit_func_charge}All fit functions for the
  charge spectra used for each target / filter combination referred to
  the set number from table~\ref{tab:cdl_lines}. Typically each line
  that is expected and visible in the data is fit. \(G\) is a normal
  Gaussian. No 'argument' to \(G\) means each parameter
  (\(N, \mu , \sigma\)) is fit. Specific arguments imply this parameter
  is fixed relative to another parameter. In both
  \(\text{Cu}-\text{EPIC}\) lines 'unknown' Gaussians are added to
  cover the behavior of the data at higher charges. It is unclear what
  the real cause is, in particular in the lower energy case.}
\end{table}
\normalsize

\footnotetext{The used EPIC (PP G12) filter refers to a filter
  developed for the EPIC camera of the XMM-Newton telescope. It is a
  bilayer of \(\SI{1600}{\textup{\r{A}}}\) polyimide and \(\SI{800}{\textup{\r{A}}} \)
  aluminium. For more information on the EPIC filters, see
  \cite{struder2001xmm_pnccd,turner2001xmm_mos,barbera2003monitoring,barbera2016thin}.}

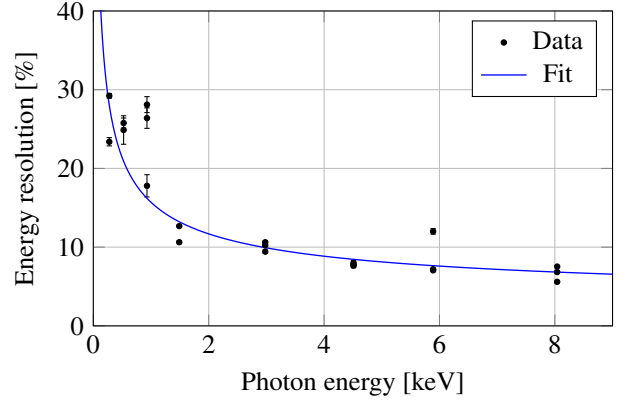
\begin{figure}[htb]
\centering
\begin{tikzpicture}
    \begin{axis}[
                  grid = major,
                  xlabel={Photon energy [keV]},
                  ylabel={Energy resolution [\%]},
				   xmin=0, xmax=9,
				   ymin = 0, ymax = 40,
				   scaled y ticks=false,
				   scaled x ticks=false,
				   width=0.95\linewidth,
				   height=0.65\linewidth,
                  legend style={at={(0.97,0.97)},anchor=north east,cells={align=left}}
                ]
        \addplot[black, only marks, mark size=1pt, error bars/.cd, y dir=both, y explicit] table[skip first n=0, x index=1, y expr=\thisrowno{2}*100, y error expr=\thisrowno{3}*100, col sep=comma] {energyresoplot-2019.csv};
        \addlegendentry{Data}
		\addplot [color = blue, line width=0.5pt, mark = none, domain=0:9, samples={600}]{(1/(7.3*sqrt(x))+0.02)*100};
        \addlegendentry{Fit}

		\end{axis}
\end{tikzpicture}
\caption{Energy resolution obtained from the CDL measurements. The
  data follows roughly the expected $1/\sqrt{E}$ behaviour indicated in blue,
  albeit larger discrepancies can be observed at low energies. Also
  seen from the multiple data sets taken for some energies a large
  spread can be observed not reflected by the error bars.}
\label{fig:CDL_energy_res}
\end{figure}

Figure~\ref{fig:cdl:calibrated_energy_histos} shows normalized
histograms of all CDL data after applying basic cuts and performing
said energy calibrations. From this the energy resolution
$\varepsilon _\text{E}$ of the detector can be extracted via
\begin{align}
\varepsilon _\text{E} = \frac{\delta E}{E}\, ,
\end{align}
with $E$ as the energy and $\delta E$ as the width of the gauss
fit. This is shown in figure~\ref{fig:CDL_energy_res} roughly following
the expected $1/\sqrt{E}$ behaviour indicated in blue.

\section{Data taking campaign}
\label{sec:cast}
The detector was installed in October 2017 at the CAST experiment and
took data until March 2018. After a maintenance break the detector was
reinstalled in October 2018 and continued until the end of 2018 with
data taking~\cite{altenmuller2025search}. In total $\SI{3516}{\hour}$ of background data were taken
during those two data taking periods with an active time of the
detector of $\SI{3157}{\hour}$.  This is a very good ratio given the
limitations of the readout chain. A breakdown of the different periods is shown in
table~\ref{tab:cast:total_data_time}.

\begin{table}[htbp]
  \centering
  \begin{tabular}{lrrr}
    \hline
          & Background [h] & Active b. [h] & Active [\%] \\[0pt]
    \hline
    Run-1 & 2391.16        & 2144.12       & 89.67       \\[0pt]
    Run-2 & 1124.93        & 1012.68       & 90.02       \\[0pt]
    Total & 3516.09        & 3157.35       & 89.80       \\[0pt]
    \hline
  \end{tabular}
  \caption{\label{tab:cast:total_data_time}Background data taken with
    the seven-GridPix detector at CAST in the time between October
    2017 and December 2018. 'Active b.' refers to the total background
    time excluding the dead time due to readout of the GridPix
    module.}
\end{table}

During the operation, the detector did perform well overall, however a
few minor issues were observed. First, the gas gain of the detector
showed some significant changes. Although not understood in all
details, it is correlated with detector temperature.

A second issue was that the FADC picked up noise. During the first
months of data taking, the FADC amplifier settings were changed twice to
guarantee stable operation. As a result, the minimum activation
threshold for the FADC---to act as a readout trigger---increased from
about $\SI{1.2}{keV}$ to $\SI{2.1}{keV}$.

The last issue was that due to a signal processing bug, the
scintillators did not work for the data taking in 2017 / beginning
2018. In the last part of the data taking, all systems operated
nominally. In total, the detector was extensively tested and the
influence of each detector feature on the background rate was
investigated.

\section{Measured background rates}
\label{sec:background}
The intended application of this detector for axion searches
requires, by nature of the experiment, very low background rates. To
achieve such background rates, special hardware features as well as
sophisticated software features are required.

The main hardware features beyond the general ability to achieve
single electron efficiency and thus reconstruct the geometric event
shapes and their energies are:
\begin{itemize}
\item scintillators allowing to tag likely background induced X-rays.
\item an FADC reading out the induced signal on the grid to act as a
trigger and provide longitudinal shape information.
\item multiple GridPixes around the central GridPix to clarify event
shapes of either sparse events or events near the edge of the chip.
\end{itemize}

Each of these requires specific methods to achieve a background
reduction. In the following, each aspect will be briefly explained and
its possible effect on the background level will be evaluated. For an
extended explanation see \cite{SchmidtPhD}. In the following
sections, after a brief introduction to the analysis framework
(section~\ref{sec:det:analysis}), in
section~\ref{sec:det:analysis} the likelihood cut method as the base
classifier will be explained. Then in
sections~\ref{sec:background:scintis}~to~\ref{sec:background:line} the
different additional vetoes will be explained, before in
section~\ref{sec:background:combined} finally the achieved background
rate is discussed.

\subsection{Analysis software and data processing}
\label{sec:det:analysis}

The data reconstruction and analysis of GridPix data is performed with
the TimepixAnalysis \cite{TPA} framework.  It handles the entire data
processing explained below. After parsing the raw data, a cluster
finding algorithm identifies individual clusters.  Each cluster is
geometrically reconstructed.
Further, each cluster is calibrated in energy, based on the
recorded charge in each pixel.
These properties serve as the basis for the initial
data classification. Each chip and cluster of the seven GridPixes is
treated separately.

\subsection{Likelihood cut method as primary data classifier}
\label{sec:background:like}

For comparison of how the new detector features
improve the background rate, the primary classifier used is a
likelihood cut-based method. The method is almost identical to the one
used in \cite{krieger2018energy,krieger2018search}, with a few minor
improvements.

Based on the X-ray tube data presented, for each energy probability
density functions $\mathcal{P} _{x}(x)$ for three geometric variables
are constructed:
\begin{enumerate}
    \item the eccentricity $\epsilon$ of the cluster, determined by computing the long and short axis of the two-dimensional cluster and then computing the ratio of the RMS of the projected positions of all active pixels within the cluster along each axis.
    \item the fraction of all pixels within a circle of the radius of one transverse RMS from the cluster centre, $f$.
    \item the length of the cluster (full extension along the long axis) divided by the transverse RMS, $l$.
\end{enumerate}
A likelihood function is defined as the product of these:
\begin{equation}\label{eq:background:likelihood_def}
\mathcal{L}(\epsilon , f, l) = \mathcal{P} _{\epsilon }(\epsilon ) \cdot \mathcal{P} _{f}(f)
\cdot \mathcal{P} _l(l)\,.
\end{equation}

To classify events as signal or background, one then sets a desired
”software efficiency” $\epsilon _{\text{eff}}$, which is defined as:
\begin{equation}\label{eq:background:lnL:cut_condition}
\epsilon _{\text{eff}} = \frac{\int _0^{\mathcal{L'}} \mathfrak{L}(\mathcal{L}) \, \mathrm{d}\mathcal{L}}{\int _0^{\infty}\mathfrak{L}(\mathcal{L}) \, \mathrm{d} \mathcal{L}},
\end{equation}
where $\mathcal{L'}$ denotes a specific $\mathfrak{L}$ value which
yields the desired efficiency. For the software efficiency
$\epsilon _{\text{eff}} = 80\,\%$ is chosen in the following, independent of the cluster energy.

For more details on this approach, see \cite{SchmidtPhD}.

\subsection{Scintillators as vetoes}
\label{sec:background:scintis}

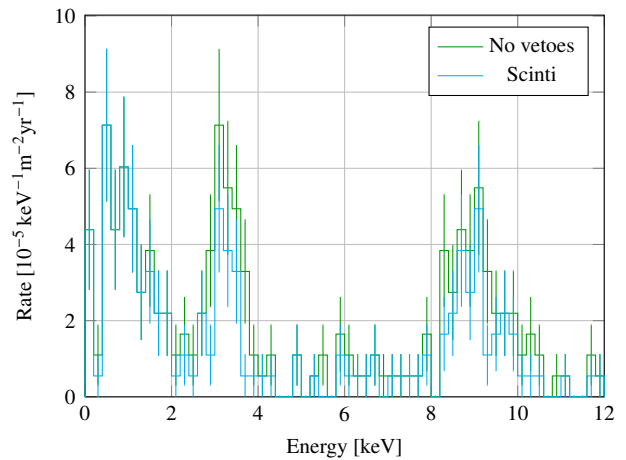
\begin{figure}[htb]
    \centering
    \begin{tikzpicture}[font=\footnotesize]
	    \begin{axis}[
            grid = major,
            xlabel={Energy [keV]},
            ylabel={Rate $[10^{-5}\,\text{keV}^{-1}\text{m}^{-2}\text{yr}^{-1}]$},
            ytick scale label code/.code={},
			xmin=0, xmax=12,
			ymin = 0,
			scaled x ticks=false,
			width=0.95\linewidth,
			height=0.75\linewidth,
            legend style={at={(0.97,0.97)},anchor=north east,cells={align=left}}
			]

        \addplot[green!60!black, ybar interval, const plot] table[skip first n=0, x index=0, y index=1, col sep=space] {./background_rate_crGold_run3.csv};
        \addplot[green!60!black, only marks, mark size=0pt, forget plot, error bars/.cd, y dir=both, y explicit] table[skip first n=0, x expr =(\thisrowno{0}+0.1), y expr =(\thisrowno{1}),y error index=2, col sep=space] {./background_rate_crGold_run3.csv};
        \addlegendentry{No vetoes}

        \addplot[cyan,ybar interval,const plot] table[skip first n=0, x index=0, y index=3, col sep=space] {./background_rate_crGold_run3.csv};
        \addplot[cyan, only marks, mark size=0pt, forget plot, error bars/.cd, y dir=both, y explicit] table[skip first n=0, x expr =(\thisrowno{0}+0.1), y expr =(\thisrowno{3}),y error index=4, col sep=space] {./background_rate_crGold_run3.csv};
        \addlegendentry{Scinti}

	\end{axis}
    \end{tikzpicture}
  \caption{\label{fig:background:scinty} Effect of the scintillator veto on the background rate in the centre
    $5 \times 5\,\text{mm}^2$ region using the
    Run-2 dataset from CAST. }
\end{figure}

The scintillators act literally as a veto. If a scintillator trigger
happened \(\leq \SI{3.75}{\micro s}\) before the FADC triggered a
readout, the corresponding event is vetoed.  In
figure~\ref{fig:background:scinty} the effect of the scintillator veto
on the background is shown for only the Run-2 dataset, since in Run-1
the veto malfunctioned. The veto shows significant improvements in the
energy region above $\sim 2\,\text{keV}$. This is expected since the
FADC needs to trigger to read out the scintillator veto. Overall, the
veto reduces the background to $\sim 79\,\%$ in the energy range from
0 to $8\,\text{keV}$. However, for the full dataset this drops since
in the first dataset no veto is available. The results of the analysis
of the full dataset are shown in table~\ref{tab:background:vetoes}. As
expected, the efficiency is only $1/3$ since it only affects $1/3$ of
the total data.  The resulting improvement of the overall background
rate is shown in cyan in figure~\ref{fig:background:comparison} using
both scintillators.

\subsection{FADC as a veto}
\label{sec:background:fadc}

\begin{figure}[htb]
\centering
\begin{tikzpicture}[font=\footnotesize]
    \begin{axis}[
                grid = major,
                xlabel={Rise time [ns]},
                ylabel=Density,
				xmin=0, xmax=400,
			    ymin=0, ymax=0.07,
				scaled y ticks=false,
				scaled x ticks=false,
                tick label style={/pgf/number format/fixed},
				width=0.95\linewidth,
				height=0.6\linewidth,
				legend style={at={(0.97,0.97)},anchor=north east,cells={align=left}}
                ]
    \addplot[green!60!black,ybar interval, draw=none, fill, opacity=0.4, area legend] table[skip first n=1, x expr=\thisrowno{0}, y index=1, col sep=space] {fadc_data_kde_riseTime_run3.csv};
    \addlegendentry{Background data}
    
	\addplot[purple!80!black, ybar interval, draw=none, fill, opacity=0.4, area legend] table[skip first n=1, x expr=\thisrowno{2}, y index=3, col sep=space] {fadc_data_kde_riseTime_run3.csv};
    \addlegendentry{${}^{55}\text{Fe}$ data}
	\end{axis}
\end{tikzpicture}
\caption{KDE of the rise time of the FADC signals in the ${}^{55}\text{Fe}$ and background data of the CAST Run-3 dataset. The X-ray data is a single peak with a mean of about $\SI{55}{ns}$ while the background distribution is extremely wide, motivating a veto based on this data.}
\label{fig:background:fadc_rise_time}
\end{figure}
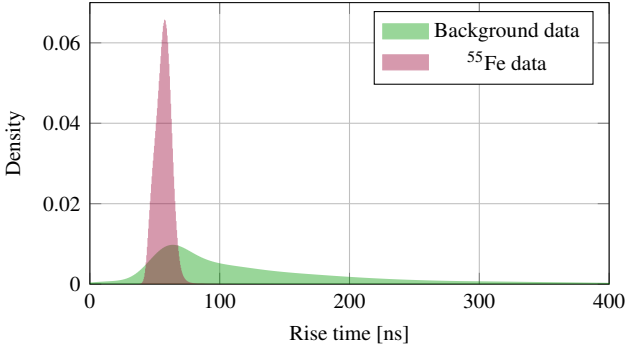

\begin{figure}[htb]
    \centering
    \begin{tikzpicture}[font=\footnotesize]
	    \begin{axis}[
            grid = major,
            xlabel={Energy [keV]},
            ylabel={Rate $[10^{-5}\,\text{keV}^{-1}\text{m}^{-2}\text{yr}^{-1}]$},
            ytick scale label code/.code={},
			xmin=0, xmax=12,
			ymin = 0,
			scaled x ticks=false,
			width=0.95\linewidth,
			height=0.75\linewidth,
            legend style={at={(0.97,0.97)},anchor=north east,cells={align=left}}
			]
        \addplot[green!60!black,ybar interval,const plot] table[skip first n=0, x index=0, y index=1, col sep=space] {./background_rate_crGold.csv};
        \addplot[green!60!black, only marks, mark size=0pt, forget plot, error bars/.cd, y dir=both, y explicit] table[skip first n=0, x expr =(\thisrowno{0}+0.1), y expr =(\thisrowno{1}),y error index=2, col sep=space] {./background_rate_crGold.csv};
        \addlegendentry{No vetoes}

        \addplot[blue,ybar interval,const plot] table[skip first n=0, x index=0, y index=5, col sep=space] {./background_rate_crGold.csv};
        \addplot[blue, only marks, mark size=0pt, forget plot, error bars/.cd, y dir=both, y explicit] table[skip first n=0, x expr =(\thisrowno{0}+0.1), y expr =(\thisrowno{5}),y error index=6, col sep=space] {./background_rate_crGold.csv};
        \addlegendentry{FADC}

	\end{axis}
    \end{tikzpicture}
  \caption{\label{fig:background:FADC} Effect of the FADC veto on the background rate in the centre
    $5 \times 5\,\text{mm}^2$ region using the
    full 2017/18 seven-GridPix detector dataset from CAST. }
\end{figure}
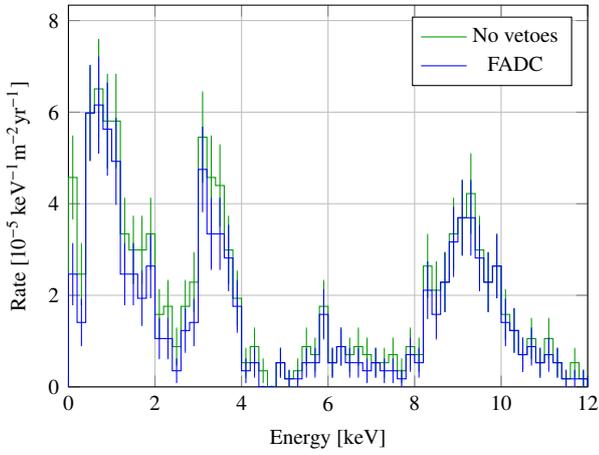
The FADC acts as a veto based on the analogue grid signals. X-rays and
orthogonal track-like events are expected to significantly differ in
their signal rise time. We perform a cut on the FADC signal rise
times, which correspond to a two-sided cut such that anything outside
the \(1^{\text{st}}\) and \(99^{\text{th}}\) percentiles of the
${}^{55}\text{Fe}$ calibration data distribution are
removed. Figure~\ref{fig:background:fadc_rise_time} shows the FADC
rise time of the calibration data against background data. By cutting
on these, the background is reduced, see
figure~\ref{fig:background:FADC}. It can be seen that an improvement
was achieved over the full energy range.

The resulting overall improvements are shown in
figure~\ref{fig:background:comparison} in blue.

\subsection{Outer GridPix as veto - 'septem veto'}
\label{sec:background:septem}
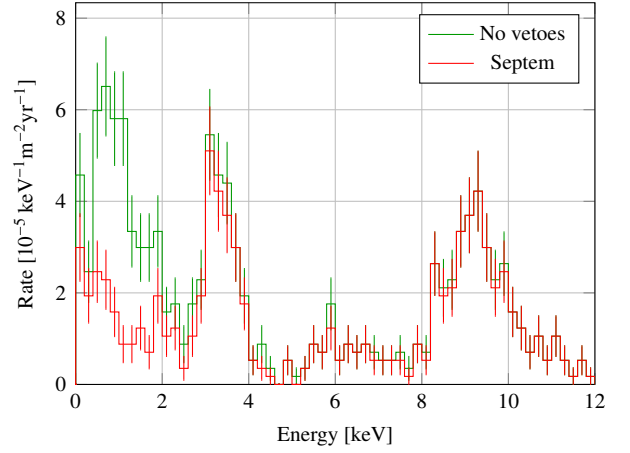
\begin{figure}[htb]
    \centering
    \begin{tikzpicture}[font=\footnotesize]
	    \begin{axis}[
            grid = major,
            xlabel={Energy [keV]},
            ylabel={Rate $[10^{-5}\,\text{keV}^{-1}\text{m}^{-2}\text{yr}^{-1}]$},
            ytick scale label code/.code={},
			xmin=0, xmax=12,
			ymin = 0,
			scaled x ticks=false,
			width=0.95\linewidth,
			height=0.75\linewidth,
            legend style={at={(0.97,0.97)},anchor=north east,cells={align=left}}
			]
        \addplot[green!60!black,ybar interval,const plot] table[skip first n=0, x index=0, y index=1, col sep=space] {./background_rate_crGold.csv};
        \addplot[green!60!black, only marks, mark size=0pt, forget plot, error bars/.cd, y dir=both, y explicit] table[skip first n=0, x expr =(\thisrowno{0}+0.1), y expr =(\thisrowno{1}),y error index=2, col sep=space] {./background_rate_crGold.csv};
        \addlegendentry{No vetoes}

        \addplot[red,ybar interval,const plot] table[skip first n=0, x index=0, y index=7, col sep=space] {./background_rate_crGold.csv};
        \addplot[red, only marks, mark size=0pt, forget plot, error bars/.cd, y dir=both, y explicit] table[skip first n=0, x expr =(\thisrowno{0}+0.1), y expr =(\thisrowno{7}),y error index=8, col sep=space] {./background_rate_crGold.csv};
        \addlegendentry{Septem}

	\end{axis}
    \end{tikzpicture}
  \caption{\label{fig:background:septem} Effect of the septem veto on the background rate in the centre
    $5 \times 5\,\text{mm}^2$ region using the
    full 2017/18 seven-GridPix detector dataset from CAST. }
\end{figure}
The six GridPixes surrounding the central GridPix can be used to
further reduce the amount of background. Depending on their location,
events may be truncated, which is especially the case for tracks
closer to the edges and corners of the GridPixes.

Therefore, for any cluster passing the likelihood cuts and other
vetoes, the outer GridPixes are checked for signals. The cluster may
thus end up being reconstructed as part of a larger cluster extending
to another chip. In this case the initial cluster is vetoed as shown
in figure~\ref{fig:septem_veto_example}. The figure shows that the
purple cluster, originally reconstructed as X-ray-like, is part of a
larger cluster extending to the outer GridPixes.

The effect of the septem veto individually is shown in
figure~\ref{fig:background:septem}. It can easily be observed that the
septem veto mainly acts on events of low energy. This is expected
since it cuts on truncated clusters, thus, only a small part of the
clusters is detected on the central GridPix.  The background rate with
the septem veto is shown in figure~\ref{fig:background:comparison} in
red, where we again see that most of the improvement is in the lower energy
range \(< \SI{2}{keV}\). This is the most important region for the
solar axion flux produced through the axion-electron coupling.

\subsection{'Line veto'}
\label{sec:background:line}
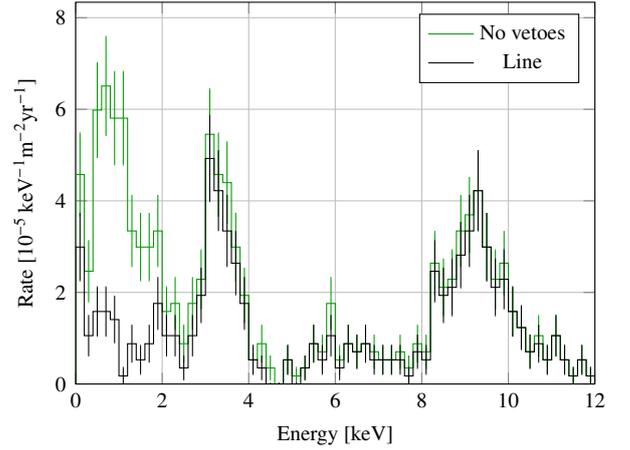
\begin{figure}[htb]
    \centering
    \begin{tikzpicture}[font=\footnotesize]
	    \begin{axis}[
            grid = major,
            xlabel={Energy [keV]},
            ylabel={Rate $[10^{-5}\,\text{keV}^{-1}\text{m}^{-2}\text{yr}^{-1}]$},
            ytick scale label code/.code={},
			xmin=0, xmax=12,
			ymin = 0,
			scaled x ticks=false,
			width=0.95\linewidth,
			height=0.75\linewidth,
            legend style={at={(0.97,0.97)},anchor=north east,cells={align=left}}
			]
        \addplot[green!60!black,ybar interval,const plot] table[skip first n=0, x index=0, y index=1, col sep=space] {./background_rate_crGold.csv};
        \addplot[green!60!black, only marks, mark size=0pt, forget plot, error bars/.cd, y dir=both, y explicit] table[skip first n=0, x expr =(\thisrowno{0}+0.1), y expr =(\thisrowno{1}),y error index=2, col sep=space] {./background_rate_crGold.csv};
        \addlegendentry{No vetoes}

        \addplot[black,ybar interval,const plot] table[skip first n=0, x index=0, y index=9, col sep=space] {./background_rate_crGold.csv};
        \addplot[black, only marks, mark size=0pt, forget plot, error bars/.cd, y dir=both, y explicit] table[skip first n=0, x expr =(\thisrowno{0}+0.1), y expr =(\thisrowno{9}),y error index=10, col sep=space] {./background_rate_crGold.csv};
        \addlegendentry{Line}

	\end{axis}
    \end{tikzpicture}
  \caption{\label{fig:background:line} Effect of the line veto on the background rate in the centre
    $5 \times 5\,\text{mm}^2$ region using the
    full 2017/18 seven-GridPix detector dataset from CAST. }
\end{figure}
\begin{figure}[htb]
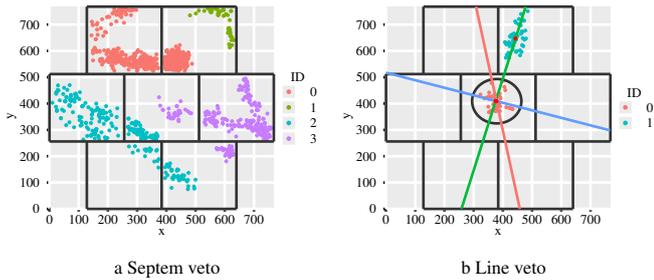

  \centering
  \begin{subfigure}{.5\linewidth}
    \centering
    \captionsetup{width=.9\linewidth}%
    

    \caption{Septem veto}
    \label{fig:septem_veto_example}
  \end{subfigure}%
  \begin{subfigure}{.5\linewidth}
    \centering
    \captionsetup{width=.9\linewidth}%
    
%

    \caption{Line veto}
    \label{fig:line_veto_example}
  \end{subfigure}
  \caption{Examples showcasing the use case for the two different vetoes utilizing the GridPix ring.
    (a): Example of an event vetoed by the septem veto, where the purple cluster on the centre is found
    to be part of larger, background like cluster. (b): Example of an event vetoed by the line veto. The
    blue cluster at the top points at the X-ray like centre cluster.}
  \label{gridpix-fig:veto_functionality}
\end{figure}

Using the outer GridPixes another veto can be constructed, the 'line
veto', which checks whether there are clusters on the outer GridPixes
whose long axis "points at" the cluster that passed the likelihood
cut. The idea being that there is a high chance that such clusters are
correlated, especially because ionization is an inherently statistical
process. An example of an event being vetoed by the 'line veto' is
shown in figure~\ref{fig:line_veto_example}.

In figure~\ref{fig:background:line} the effect of the line veto
applied to the full uncut dataset can be seen. It is visible that
again as for the septem veto the reduction mainly affects clusters at
energies below $\sim 2\,\text{keV}$. The reduction is somewhat similar
to the septem veto, since many clusters identified by the septem veto
will be also identified by the line veto. However, due to the addition
of allowing larger gaps in between the clusters also some additional
background will be vetoed.  This can be seen in the overall background rate with the line
veto shown in black in figure~\ref{fig:background:comparison}.

\subsection{Note on the efficiency of septem \& line veto}
\label{sec:org21a83ed}
Due to the detector limitation of only triggering a readout via the
FADC for the central chip, there is an inherent possibility of random
coincidences for any clusters on the outer chips. The rate of random
coincidences was simulated to about \(\SI{16.9}{\%}\) for the septem
veto, \(\SI{14.6}{\%}\) for the line veto, and \(\SI{21.4}{\%}\) for the
combination of both. The origin of these high numbers is the fact of very long shutter times $\mathcal{O}(2\,\text{s})$ and the absence of a ToA measurement. Thus events registered only on this outer 6 GridPixes will not trigger the readout, and can therefore later be wrongly put into coincidence with a hit on the central GridPix. 

\subsection{Summarizing the vetoes}
\begin{table}[htbp]
  \centering
  \begin{tabular}{lrrr}
    \hline
    Energy      & $0\text{-}8\,\text{keV}$ & $0.5\text{-}5\,\text{keV}$ & $2\text{-}8\,\text{keV}$ \\[0pt]
    \hline
    Scintillator & 93.17\,\%    & 92.49\,\%   & 86.75\,\%   \\[0pt]
    FADC         & 78.88\,\%    & 80.35\,\%   & 75.21\,\%   \\[0pt]
    septem       & 60.87\,\%    & 56.65\,\%   & 84.62\,\%   \\[0pt]
    line         & 53.83\,\%    & 50.87\,\%   & 79.92\,\%   \\[0pt]
    \hline
  \end{tabular}
  \caption{\label{tab:background:vetoes}Percental reduction of the
    background using only singular vetoes implemented in the detector
    for the full background data. Three different energy ranges are
    selected. $0\text{-}8\,\text{keV}$ for the overall background,
    $0.5\text{-}5\,\text{keV}$ for the background range interesting
    for the axion-electron coupling, and $2\text{-}8\,\text{keV}$ as
    the interesting range for the axion-photon coupling.}
\end{table}

To summarize, the effects on the background level of all vetoes are
shown in table~\ref{tab:background:vetoes} for three different energy
ranges. The first energy range $0\text{-}8\,\text{keV}$ concludes the
overall detector performance for solar axions. Here, it can be seen
that all vetoes perform well, albeit the septem and the line veto
improve the background the most. However, it has to be stated that by
using all vetoes combined, an even better background rejection is
achieved as shown in figure~\ref{fig:background:comparison}. The other
two regions are more specific to either axion-electron-coupling
($0.5\text{-}5\,\text{keV}$) or axion-photon coupling
($2\text{-}8\,\text{keV}$). It can be seen that for axion-photon
coupling the FADC veto adds the largest contribution to the background
reduction, as also the scintillator veto would have, if it had been
active for the full measurement time. This comes from the fact that
both the septem and the line veto mostly reduce the background at very
low energies.

\subsection{Final background rate using all vetoes}
\label{sec:background:combined}
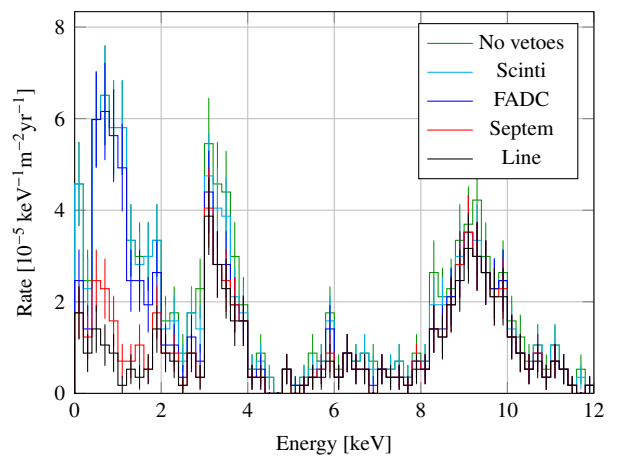
\begin{figure}[htb]
    \centering
    \begin{tikzpicture}[font=\footnotesize]
	    \begin{axis}[
            grid = major,
            xlabel={Energy [keV]},
            ylabel={Rate $[10^{-5}\,\text{keV}^{-1}\text{m}^{-2}\text{yr}^{-1}]$},
            ytick scale label code/.code={},
			xmin=0, xmax=12,
			ymin = 0,
			scaled x ticks=false,
			width=0.95\linewidth,
			height=0.75\linewidth,
            legend style={at={(0.97,0.97)},anchor=north east,cells={align=left}}
			]
        \addplot[green!60!black,ybar interval,const plot] table[skip first n=0, x index=0, y index=1, col sep=space] {./CAST_BG_LnL_vetos.csv};
        \addplot[green!60!black, only marks, mark size=0pt, forget plot, error bars/.cd, y dir=both, y explicit] table[skip first n=0, x expr =(\thisrowno{0}+0.1), y expr =(\thisrowno{1}),y error index=2, col sep=space] {./CAST_BG_LnL_vetos.csv};
        \addlegendentry{No vetoes}

        \addplot[cyan,ybar interval,const plot] table[skip first n=0, x index=0, y index=3, col sep=space] {./CAST_BG_LnL_vetos.csv};
        \addplot[cyan, only marks, mark size=0pt, forget plot, error bars/.cd, y dir=both, y explicit] table[skip first n=0, x expr =(\thisrowno{0}+0.1), y expr =(\thisrowno{3}),y error index=4, col sep=space] {./CAST_BG_LnL_vetos.csv};
        \addlegendentry{Scinti}

        \addplot[blue,ybar interval,const plot] table[skip first n=0, x index=0, y index=5, col sep=space] {./CAST_BG_LnL_vetos.csv};
        \addplot[blue, only marks, mark size=0pt, forget plot, error bars/.cd, y dir=both, y explicit] table[skip first n=0, x expr =(\thisrowno{0}+0.1), y expr =(\thisrowno{5}),y error index=6, col sep=space] {./CAST_BG_LnL_vetos.csv};
        \addlegendentry{FADC}

        \addplot[red,ybar interval,const plot] table[skip first n=0, x index=0, y index=7, col sep=space] {./CAST_BG_LnL_vetos.csv};
        \addplot[red, only marks, mark size=0pt, forget plot, error bars/.cd, y dir=both, y explicit] table[skip first n=0, x expr =(\thisrowno{0}+0.1), y expr =(\thisrowno{7}),y error index=8, col sep=space] {./CAST_BG_LnL_vetos.csv};
        \addlegendentry{Septem}

        \addplot[black,ybar interval,const plot] table[skip first n=0, x index=0, y index=9, col sep=space] {./CAST_BG_LnL_vetos.csv};
        \addplot[black, only marks, mark size=0pt, forget plot, error bars/.cd, y dir=both, y explicit] table[skip first n=0, x expr =(\thisrowno{0}+0.1), y expr =(\thisrowno{9}),y error index=10, col sep=space] {./CAST_BG_LnL_vetos.csv};
        \addlegendentry{Line}
	\end{axis}
    \end{tikzpicture}
  \caption{\label{fig:background:comparison} Background rate in the centre
    $5 \times 5\,\text{mm}^2$ region using the
    full 2017/18 seven-GridPix detector dataset from CAST. Each successive veto,
    applied in the order of the legend, is shown
    cumulatively. The 'Line veto' contains all discussed vetoes.}
\end{figure}

All the vetoes discussed above yield a very good improvement of the
background rate, shown in figure~\ref{fig:background:comparison},
which contains the comparisons of all vetoes discussed above. Each
veto builds on the previous ones (in the order as they were mentioned
above). The hardware vetoes lead to a significant reduction of the
background. Especially the FADC veto and the septem veto perform very
well reducing the background in the order of 20\,\% (FADC) and 30\,\%
(septem). While the FADC reduces the background over the whole energy
range the septem veto has the largest impact at the very low energies.

\begin{figure}[htb]
    \vspace*{-0.2cm}
    \begin{subfigure}{0.5\linewidth}
        \begin{tikzpicture}[font=\footnotesize]
            \node[anchor=south west,inner sep=0] (Bild) at (0,0) {\includegraphics[width=\linewidth]{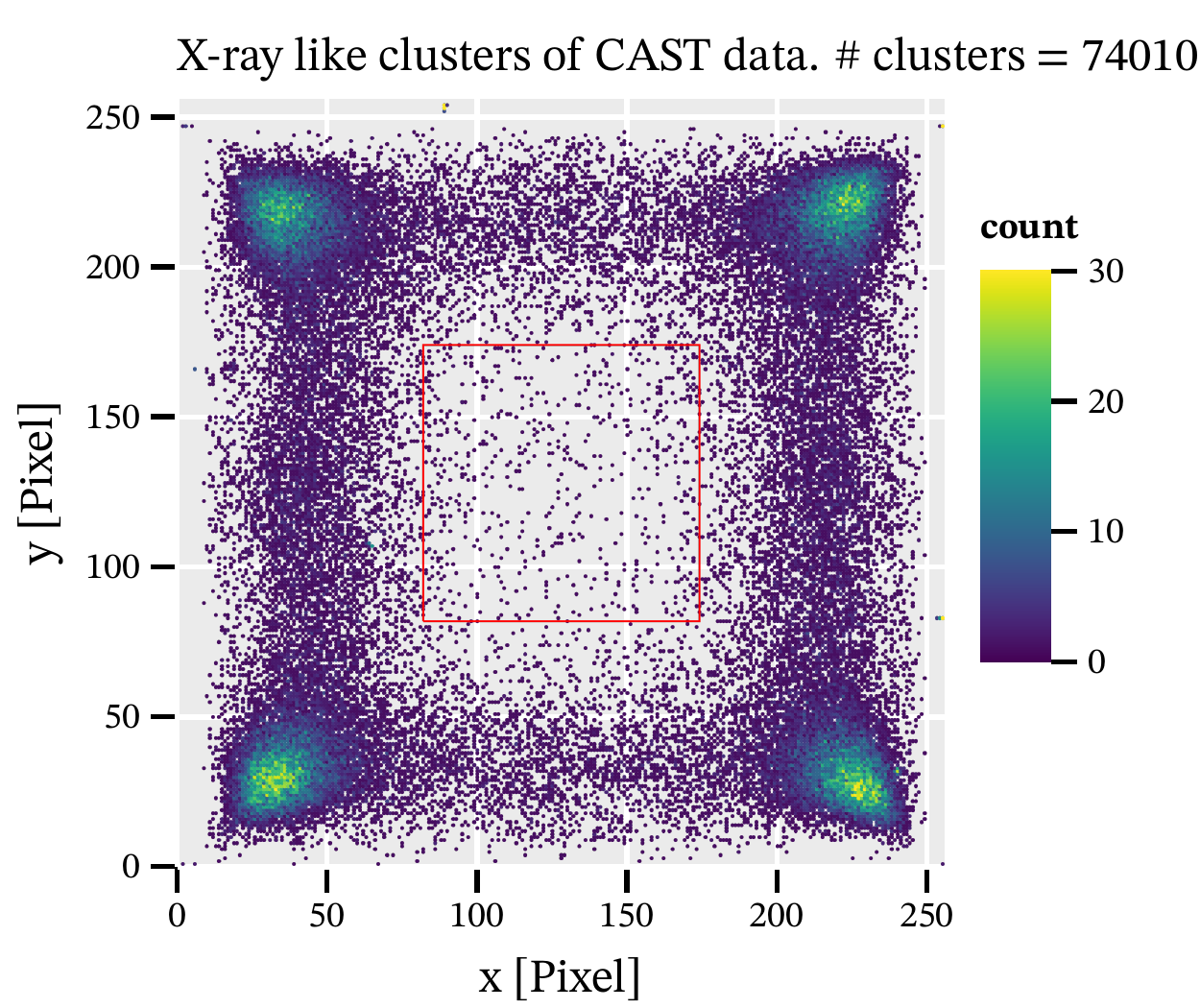}};
            \begin{scope}[x=(Bild.south east),y=(Bild.north west)]
                 \draw[draw=none,fill=white](0,1)rectangle(1,0.915);
            \end{scope}
        \end{tikzpicture}
        \caption{No vetoes}
        \label{fig:background:cluster_centers_lnL80_without}
    \end{subfigure}%
    \begin{subfigure}{0.5\linewidth}
        \begin{tikzpicture}[font=\footnotesize]
            \node[anchor=south west,inner sep=0] (Bild) at (0,0) {\includegraphics[width=\linewidth]{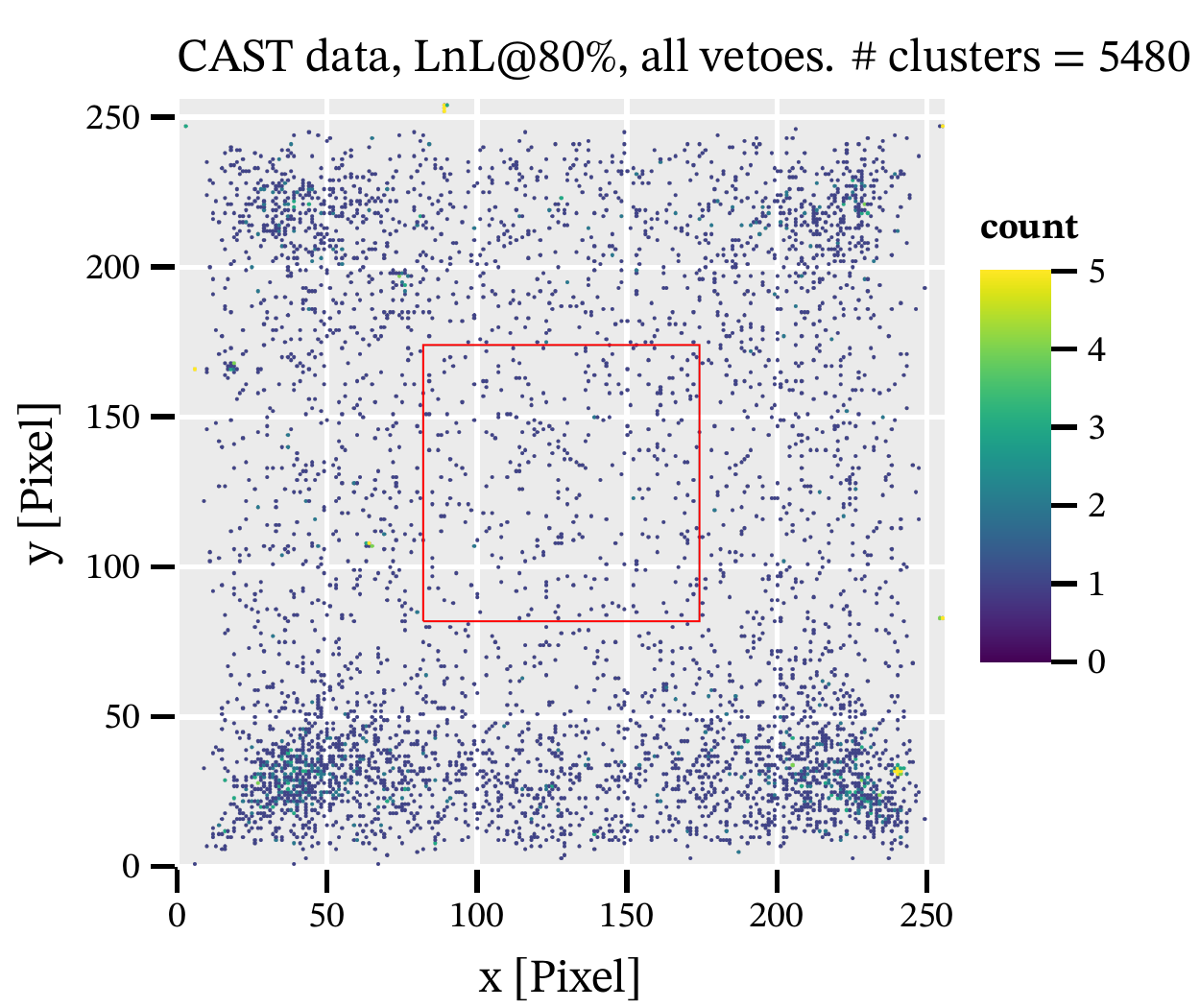}};
            \begin{scope}[x=(Bild.south east),y=(Bild.north west)]
               \draw[draw=none,fill=white](0,1)rectangle(1,0.915);
            \end{scope}
        \end{tikzpicture}
        \caption{All vetoes}
        \label{fig:background:cluster_centers_lnL80_all}
    \end{subfigure}%
    \caption{Cluster centres of all X-ray like clusters in the 2017/18
      CAST background
      data. (\subref{fig:background:cluster_centers_lnL80_without})
      shows the remaining data for the likelihood~method without any
      vetoes. (\subref{fig:background:cluster_centers_lnL80_all})
      includes all vetoes. The vetoes lead to a dramatic reduction in
      background especially outside the centre
      $\SI[parse-numbers=false]{5 \times 5}{mm^2}$ (red square).}
  \label{fig:background:cluster_center_comparison}
\end{figure}

In addition, figure~\ref{fig:background:cluster_center_comparison} shows
a comparison of all cluster centres left after only applying the $\ln\mathcal{L}$~method
at \(\epsilon _s = \SI{80}{\%}\) and adding all vetoes. The background
rejection improves massively towards the edges and corners, as one
would expect thanks to the outer GridPixes.

The background rate between \(\SIrange{0}{8}{keV}\) ends up at
\(\SI{7.9164(5901)e-06}{keV^{-1} cm^{-2} s^{-1}}\). However, note that
the combined efficiency drops to $\SI{61.6}{\%}$ due to the efficiency
penalty of the septem and line vetoes.

\section{Summary \& conclusion}
\label{sec:summary}
The seven-GridPix detector presented in this paper represents a major
improvement over previous detectors. Compared to the single GridPix
detector of \cite{krieger17_gridpix} it significantly improves the detection
efficiency at low energies, due to the much thinner silicon-nitride
window. The additional detector features yield a significant (28\,\%) step in
background rejection.

Compared to a state-of-the-art Micromegas detector
\cite{castcollaboration2024new}, this detector manages to narrow the gap
in background rate, while providing lower energy thresholds and higher
detection efficiencies, in particular below \(\SI{2}{keV}\).

Future detectors will utilize GridPix3 (based on Timepix3), which allow
for simultaneous time over threshold and time of arrival measurement
and data-driven readouts. These will make the usage of the FADC
redundant, avoiding any issues with potential noise and allowing for
the septem and line vetoes without any efficiently
penalties. Therefore, the roadmap to significant improvements from the
here introduced detector is clear.



\bibliographystyle{elsarticle-num}
\bibliography{references}





\end{document}